\documentclass[12pt]{article}

\usepackage[utf8]{inputenc}
\usepackage{textcomp}
\usepackage{chetnew}
\usepackage{tikz}
\usepackage{amsmath,amssymb,amsfonts,mathrsfs,dsfont,yfonts,bbm}\usetikzlibrary{calc}
\usepackage{xcolor}
\usepackage{braket}
\usepackage[display]{texpower}
\usepackage{graphicx}
\usepackage{relsize}

\newcommand{\be}{\begin{equation}}
\newcommand{\ee}{\end{equation}}
\newcommand{\bea}{\begin{eqnarray}}
\newcommand{\eea}{\end{eqnarray}}

\newcommand{\CL}{\mathcal{L}}
\newcommand{\CK}{\mathcal{K}}
\newcommand{\CA}{\mathcal{A}}
\newcommand{\CH}{\mathcal{H}}

\newcommand{\CC}{\mathcal{C}}

\newcommand{\CT}{\mathcal{T}}

\newcommand{\CN}{\mathcal{N}}
\newcommand{\CS}{\mathcal{S}}
\newcommand{\CM}{\mathcal{M}}

\newcommand*{\boxcoloro}{orange}
\makeatletter
\newcommand{\boxedo}[1]{\textcolor{\boxcoloro}{%
\tikz[baseline={([yshift=-1ex]current bounding box.center)}] \node [rectangle, minimum width=1ex,rounded corners,draw] {\normalcolor\m@th$\displaystyle#1$};}}
 \makeatother
\newcommand*{\boxcolorr}{red}
\makeatletter
\newcommand{\boxedr}[1]{\textcolor{\boxcolorr}{%
\tikz[baseline={([yshift=-1ex]current bounding box.center)}] \node [rectangle, minimum width=1ex,rounded corners,draw] {\normalcolor\m@th$\displaystyle#1$};}}
 \makeatother
\newcommand*{\boxcolorb}{blue}
\makeatletter
\newcommand{\boxedb}[1]{\textcolor{\boxcolorb}{%
\tikz[baseline={([yshift=-1ex]current bounding box.center)}] \node [rectangle, minimum width=1ex,rounded corners,draw] {\normalcolor\m@th$\displaystyle#1$};}}
 \makeatother
\newcommand*{\boxcolorg}{green}
\makeatletter
\newcommand{\boxedg}[1]{\textcolor{\boxcolorg}{%
\tikz[baseline={([yshift=-1ex]current bounding box.center)}] \node [rectangle, minimum width=1ex,rounded corners,draw] {\normalcolor\m@th$\displaystyle#1$};}}
 \makeatother
 \newcommand*{\boxcolorp}{purple}
\makeatletter
\newcommand{\boxedp}[1]{\textcolor{\boxcolorp}{%
\tikz[baseline={([yshift=-1ex]current bounding box.center)}] \node [rectangle, minimum width=1ex,rounded corners,draw] {\normalcolor\m@th$\displaystyle#1$};}}
 \makeatother
  \newcommand*{\boxcolorc}{cyan}
\makeatletter
\newcommand{\boxedc}[1]{\textcolor{\boxcolorc}{%
\tikz[baseline={([yshift=-1ex]current bounding box.center)}] \node [rectangle, minimum width=1ex,rounded corners,draw] {\normalcolor\m@th$\displaystyle#1$};}}
 \makeatother
  \newcommand*{\boxcolory}{yellow}
\makeatletter
\newcommand{\boxedy}[1]{\textcolor{\boxcolory}{%
\tikz[baseline={([yshift=-1ex]current bounding box.center)}] \node [rectangle, minimum width=1ex,rounded corners,draw] {\normalcolor\m@th$\displaystyle#1$};}}
 \makeatother

\usepackage{amsmath}	
\begin{document}
\preprint{QMUL-PH-19-36}

\title{Galois Conjugation and Multiboundary\\[2mm] Entanglement Entropy}

\author{Matthew Buican$^{\diamondsuit}$ and Rajath Radhakrishnan$^{\clubsuit}$}

\affiliation{\smallskip CRST and School of Physics and Astronomy\\
Queen Mary University of London, London E1 4NS
\emails{$^{\diamondsuit}$m.buican@qmul.ac.uk, $^{\clubsuit}$r.k.radhakrishnan@qmul.ac.uk}}

\abstract{We revisit certain natural algebraic transformations on the space of 3D topological quantum field theories (TQFTs) called \lq\lq Galois conjugations." Using a notion of multiboundary entanglement entropy (MEE) defined for TQFTs on compact 3-manifolds with disjoint boundaries, we give these abstract transformations additional physical meaning. In the process, we prove a theorem on the invariance of MEE along orbits of the Galois action in the case of arbitrary Abelian theories defined on any link complement in $S^3$. We then give a generalization to non-Abelian TQFTs living on certain infinite classes of torus link complements. Along the way, we find an interplay between the modular data of non-Abelian TQFTs, the topology of the ambient spacetime, and the Galois action. These results are suggestive of a deeper connection between entanglement and fusion.}

\date{December 2019}

\setcounter{tocdepth}{2}

\maketitle
\toc

\newsec{Introduction}
One successful non-perturbative approach to studying many quantum field theories (QFTs) is to impose consistency conditions following from some underlying symmetries and then use these constraints to, at least partially, solve the QFTs. Recent examples of this program include constraints arising from associativity of the operator product expansion (OPE) in the modern conformal bootstrap, constraints due to holomorphy in supersymmetric field theories, and constraints coming from integrability in various settings. 

Three-dimensional\footnote{Throughout this paper, dimension means spacetime dimension.} topological quantum field theory (TQFT) is a particularly well-suited and interesting arena in which to apply this approach. Indeed, while 3D TQFTs have trivial dynamics and are characterized by protected sets of observables, they also describe a variety of phenomena of genuine physical interest (e.g., the fractional quantum Hall effect\footnote{For now, we are being somewhat careless in not distinguishing between TQFTs and spin-TQFTs. In the remainder of this paper, we will focus exclusively on the former.} and the physics of anyons \cite{Moore:1991ks}). In this setting, the basic symmetries are associativity and braiding of non-local line operators, and the corresponding consistency conditions take the form of a set of polynomial equations called the Pentagon and Hexagon equations (combined with constraints from modularity) \cite{Moore:1988qv,moore1990lectures,bakalov2001lectures,Kitaev_2006}. This approach has yielded various interesting results constraining the space of TQFTs (e.g., see \cite{rowell2009classification,bruillard2016rank}) and promises to lead to more (e.g., see \cite{Bonderson_2019,wen2019distinguish}).

Thinking in this abstract way about 3D TQFT leads to questions more familiar to high-energy theorists in other settings. For example, is the space of (unitary) 3D TQFTs populated only by Lagrangian theories \cite{Witten:1988sy,hong2008exotic}? What are the natural symmetries and dualities that act on the space of 3D TQFTs (e.g., see recent work in \cite{Delmastro:2019vnj})? At the same time, this approach complements a more Lagrangian way of thinking, based on Chern-Simons (CS) theory, that has well-known connections with various aspects of knot theory \cite{witten1989quantum}.

In this paper, our primary goal is to better understand a natural set of algebraic transformations---called \lq\lq Galois conjugations"---that map 3D TQFTs to other 3D TQFTs and arise when we imagine these theories as corresponding to solutions of the Pentagon and Hexagon equations with appropriate modular data.\footnote{Note that these transformations are not, in general, symmetries or dualities. Instead, Galois transformations are ways to jump around in the space of TQFTs while preserving many physically interesting properties.} In this context, we will mostly avoid concrete Lagrangians and think of the TQFTs as arising from well-known algebraic objects called \lq\lq Modular Tensor Categories," (MTCs) \cite{Moore:1988qv,moore1990lectures,bakalov2001lectures,Kitaev_2006}.

While the application of Galois theory to TQFT is somewhat abstract and has therefore attracted mathematical interest \cite{davidovich2013arithmetic,etingof2005fusion}, such transformations have also played a role in the condensed matter literature \cite{freedman2012galois,Lootens:2019xjv}, and, most prominently, in the closely related 2D rational conformal field theory (RCFT) literature \cite{DeBoer:1990em,coste1994remarks,Coste:1999yc,bantay2003kernel,Harvey:2018rdc}.\footnote{The appearance of Galois actions in both 2D RCFT and 3D TQFT is to be expected from the Wess-Zumino-Witten model/CS bulk-boundary correspondence. In this context, MTCs capture braiding and fusion properties of RCFT chiral algebra primaries.} Moreover, as we will review, one well-known physical consequence of the MTC structure is that Galois transformations take one between different theories while preserving a common fusion algebra and, as a consequence, any 1-form symmetry groups.\footnote{The question of what happens to more general TQFT symmetry structures under Galois conjugation will be discussed elsewhere \cite{galsymm}.}

The purpose of this paper is to gain additional physical insight into Galois transformations that goes beyond symmetry and fusion. In particular, we will study the effects of Galois transformations on a type of \lq\lq multiboundary" topological entanglement entropy (MEE) defined in \cite{Salton:2016qpp,balasubramanian2017multi,Balasubramanian2018EntanglementEA}. MEE is quite different from the more familiar entanglement entropy studied in \cite{kitaev2006topological,Levin:2006zz}. Indeed, it involves first placing TQFTs on link complements, particular compact 3-manifolds that have multiple disjoint boundaries, and then tracing out Hilbert spaces associated with proper subsets of these boundaries. MEE is therefore highly non-local. Moreover, as we will see, MEE has interesting connections with knot theory, and we will phrase properties of Galois transformations in terms of the topology of knots and links.\footnote{See \cite{Gannon:1995th} for an early study of Galois transformations and link invariants.}

Our main claim is that, for a TQFT defined on $\CM_{\CL}$, the MEE we obtain by tracing out Hilbert subspaces associated with proper subsets of the disjoint boundaries is often invariant under the TQFT Galois action. In particular, we argue that the MEEs associated with any Abelian TQFT on any link complement in $S^3$ are invariant under the TQFT Galois action. In the case of non-abelian theories, the situation is more subtle. Building on our Abelian proof and taking into account recent results on classifications of MTCs \cite{mignard2017modular,Bonderson_2019,wen2019distinguish}, we argue that a natural place to look for Galois invariance of MEE in non-Abelian theories is on 3-manifolds corresponding to complements of torus links. Indeed, we then identify infinite sets of torus link complements that give rise to invariant MEE along Galois orbits.\footnote{In the case of non-Abelian theories, the Galois action will, in general, take unitary theories to non-unitary ones. This fact leads to subtleties when defining what we mean by MEE in these latter theories. However, it turns out that there is in fact a natural definition of MEE even in the case of non-unitary theories.} As we will see, there is an interesting interplay between the topology of these link complements and basic modular data of the non-Abelian TQFTs living on these spaces (we highlight a simple application of this result in the conclusions).

While many of the constructions we present in this paper suggest generalizations beyond the world of TQFT, we leave a discussion of Galois transformations in more general 3D QFT for future work. However, we will briefly mention such generalizations in our conclusions.

The plan of the paper is as follows. In the next section, we review basic aspects of 3D TQFTs and reformulate them in the algebraic language of MTCs. This perspective is particularly suited to our discussion of Galois conjugation in Sec. \ref{galoisint}. Our review of relevant concepts then concludes in Sec. \ref{TEE}, where we define MEE.  With these concepts under our belt, in Sec. \ref{abeliansec} we prove universal results on MEE in Abelian TQFTs. The following section is dedicated to generalizing this discussion to non-Abelian TQFTs living on torus link complements. Finally, we conclude with some comments on open problems and applications suggested by our work.

\newsec{TQFT and MTC basics}\label{TQFTbasics}
As alluded to in the introduction, large classes of 3D TQFTs have a description in terms of algebraic objects called MTCs.\footnote{Strictly speaking, any MTC gives rise to a TQFT. Given reasonable axioms, including having a finite number of simple objects, the converse is also expected to be true \cite{bakalov2001lectures,wang2010topological}.} The importance of this algebraic description of a TQFT is that it allows us to think of a TQFT as a solution to a finite number of polynomial equations rather than being tied to a Lagrangian. In particular, we will see that this perspective naturally leads to Galois theory when one thinks about the space of consistent TQFTs. 

The basic constituents of an MTC are a set of labels, $\{a,b,\cdots\}$, satisfying fusion rules\footnote{Throughout this discussion, one may think of these labels as generalizing the concept of Wilson lines in CS theory. Due to the topological nature of the theory, the fusion rules correspond to a position-independent OPE.}
\be\label{fusdef}
a \otimes b = \sum_c N_{ab}^c c~, \ \ \ N_{ab}^c\in\mathbb{Z}_{\ge0}~.
\ee
Physically, the labels are the charges of the quasiparticles in the topological phase. The fusion rules provide information on how the particles can combine to form new ones. 
Indeed, the $N_{ab}^c$ coefficients measure the number of ways in which $a$ and $b$ can combine to form $c$, and $a \otimes b = c$ is allowed if and only if $N_{ab}^c>0$. In particular, the $N_{ab}^c$ are dimensions of Hilbert spaces, $V_{ab}^c$, called \lq\lq fusion spaces." More generally, the fusion space corresponding to $a_1 \otimes a_2\cdots \otimes a_n=b$ is denoted as $V_{a_1a_2\cdots a_n}^b$. As a bit of terminology, we call an MTC abelian if each fusion outcome in \eqref{fusdef} is unique (and therefore, after imposing commutativity, has fusion rules given by a finite abelian group). Otherwise, the theory is called non-abelian.

\tikzset{every picture/.style={line width=0.75pt}} 

\begin{figure}[h!]
\centering
\begin{tikzpicture}[x=0.75pt,y=0.75pt,yscale=-0.8,xscale=0.8]

\draw    (372.17,82.33) -- (414.17,140.33) ;

\draw    (414.17,140.33) -- (414.17,177.33) -- (414.17,182.33) ;

\draw    (165.67,142.83) -- (165.67,163.33) -- (165.67,181.83) ;

\draw    (165.67,142.83) .. controls (155.17,136.33) and (127.17,110.33) .. (198.17,82.33) ;

\draw    (137.17,80.33) .. controls (144.17,85.33) and (153.17,93.33) .. (159.17,97.33) ;

\draw    (169.17,104.33) .. controls (187.17,112.33) and (193.17,128.33) .. (165.67,142.83) ;

\draw    (459.17,82.33) -- (414.17,140.33) ;

\draw (137,69) node   {$a$};
\draw (373,71) node   {$a$};
\draw (461,71) node   {$b$};
\draw (201,71) node   {$b$};
\draw (166,190) node   {$c$};
\draw (415,189) node   {$c$};
\draw (290,136) node   {$=\ \ \ R^{c}_{ab}$};
\end{tikzpicture}
\caption{Action of the R-matrix}\label{Rmat}
\end{figure}
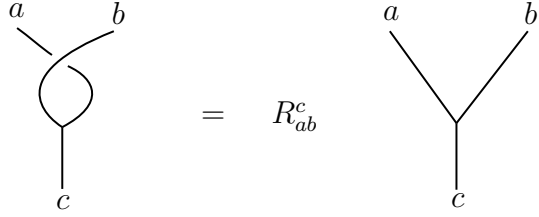

One can in principle start with any set of labels and fusion rules. However, a consistent MTC exists only if a set of consistency conditions called the Pentagon and Hexagon equations are satisfied \cite{Moore:1988qv,moore1990lectures,bakalov2001lectures,Kitaev_2006}. These relations arise due to the commutativity and associativity of the fusion operation. Commutativity of the fusion operation implies the existence of an isomorphism $V_{ab}^c \cong V_{ba}^c$. The linear map corresponding to this isomorphism is called the $R$ matrix (see Fig. \ref{Rmat}). Using the $R$ matrix, one can braid the quasiparticles around one another, and one generally finds anyonic statistics. Hence, the labels we have introduced above are often referred to as \lq\lq anyons" in the physics literature.\footnote{Performing this braiding with Wilson lines in generic Chern-Simons theories also shows that Wilson lines are typically anyonic objects.} 

\tikzset{every picture/.style={line width=0.75pt}} 

\begin{figure}[h!]
\centering
\begin{tikzpicture}[x=0.75pt,y=0.75pt,yscale=-0.8,xscale=0.8]

\draw    (491.17,81.33) -- (431.17,147.33) ;

\draw    (368.17,80.33) -- (431.17,147.33) ;

\draw    (429.17,80.33) -- (461.17,114.33) ;

\draw    (431.17,147.33) -- (431.17,186.33) ;

\draw    (218.17,80.33) -- (164.17,146.33) ;

\draw    (101.17,79.33) -- (164.17,146.33) ;

\draw    (159.17,80.33) -- (132.67,112.83) ;

\draw    (164.17,146.33) -- (164.17,185.33) ;

\draw (101,70) node   {$a$};
\draw (370,70) node   {$a$};
\draw (432,70) node   {$b$};
\draw (492,71) node   {$c$};
\draw (161,70) node   {$b$};
\draw (220,71) node   {$c$};
\draw (164,193) node   {$d$};
\draw (431,194) node   {$d$};
\draw (140,133) node   {$e$};
\draw (295,135) node   {$=\sum _{f}\left( F^{d}_{abc}\right)^{f}_{e}$};
\draw (456,133) node   {$f$};
\end{tikzpicture}
\caption{Action of the F-matrix}
\label{fusionmatrix}
\end{figure}
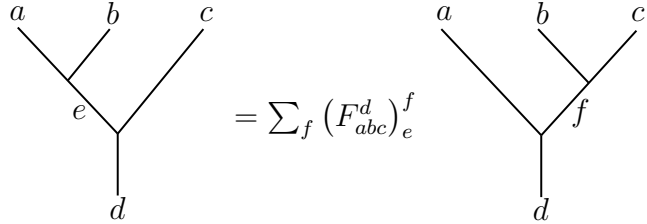

On the other hand, the associativity of the fusion operation implies that the fusion space $V_{abc}^d= \sum_e V_{ab}^e \otimes V_{ec}^d$ can also be decomposed as $V_{abc}^d=\sum_f V_{bc}^f \otimes V_{fa}^d$. The linear map corresponding to the isomorphism $\sum_e V_{ab}^e \otimes V_{ec}^d \cong \sum_f V_{bc}^f \otimes V_{fa}^d$ is called the $F$ matrix (see Fig. \ref{fusionmatrix}). As a result, we have 
\be
R_{ab}^c: V_{ab}^c \rightarrow V_{ba}^c~, \ \ \  F_{abc}^d: \sum_e V_{ab}^e \otimes V_{ec}^d  \rightarrow \sum_f V_{bc}^f \otimes V_{fa}^d~.
\ee
From the action of the $F$ matrices on $V_{abcd}^e$, one can deduce that they must satisfy the following constraint
\be
\label{pentagon}
\big (F_{a,b,k}^{e}\big)_l^i\big(F_{i,c,d}^{e}\big)_j^k=\sum_m \big(F_{b,c,d}^l\big)_m^k\big(F_{a,m,d}^{e}\big)_j^l\big(F_{a,b,c}^j\big)_i^m~.
\ee
This is the Pentagon equation. Also, from the action of the $R$ and $F$ matrices on $V_{abc}^d$, one can see that they must satisfy the Hexagon equations
\begin{eqnarray}\label{hexagon}
R_{a,c}^k\big(F_{b,a,c}^{d}\big)_i^kR_{a,b}^i&=&\sum_j \big(F_{b,c,a}^{d}\big)_j^kR_{a,j}^{d}\big(F_{a,b,c}^d\big)_i^j~,\cr
R_{c,a}^k\Big (\big(F_{b,a,c}^{d}\big)_i^k\Big )^{-1 }R_{b,a}^i&=&\sum_j \Big (\big(F_{b,c,a}^{d}\big)_j^k\Big )^{-1}R_{j,a}^{d}\Big(\big(F_{a,b,c}^d\big)_i^j\Big )^{-1}~.
\end{eqnarray}
These are polynomial equations, and they can have at most a finite number of inequivalent solutions \cite{etingof2005fusion,etingof2016tensor}.

In addition, we demand that these equations give rise to a unitary (projective) representation of the modular group,\footnote{The unitarity of this representation does not imply unitarity of the TQFT.} $SL(2,\mathds{Z})$, via
\be\label{moddata}
T_{aa}=d_a^{-1} \sum_c d_c R_{aa}^c=\theta(a)~, \ \ \  S_{a,b}=\frac{1}{\sqrt{\sum_e d_e}} \sum_c d_c \text{Tr}(R_{ab}^cR^c_{ba})~,
\ee
where $d_a$ is the $S^3$ link invariant of an unknot labelled by $a$ called the \lq\lq quantum dimension."\footnote{Related equations arise for the \lq\lq punctured" S-matrix.} It is straightforward to check that the matrices in \eqref{moddata} form a projective representation of $SL(2,\mathds{Z})$
\be\label{sl2z}
(ST)^3=\Theta C~, \ \ \ S^2=C~, \ \ \  C^2=I~,
\ee
where $\Theta=\frac{1}{\sqrt{\sum_c d_c^2}} \sum_a d_a^2 T_{aa}$, and $C$ is the charge conjugation matrix. Note that given solutions $(S, T, \Theta)$ to the above equations, we can also consider solutions with $(S,T,\Theta)\to(-S,T,-\Theta)$. Indeed, such transformations naturally arise when considering certain non-unitary MTCs like the Lee-Yang theory. Apart from certain particular cases, we will mostly consider the $S$ matrix of \eqref{moddata} in the explicit examples below. However, our results apply to both sign choices in principle.

Let us conclude our review of the modular properties of MTCs by recalling that the fusion coefficients, $N_{ab}^c$, are famously determined by the $S$ matrix elements through the Verlinde formula
\be
\label{Verlinde}
N_{ab}^c= \sum_e \frac{S_{ae}S_{be}S_{ec^*}}{S_{0e}}~.
\ee

The set of solutions to \eqref{pentagon} and \eqref{hexagon} admits a cohomological interpretation: for example, in the case of abelian MTCs, $(F,R)$ take values in abelian group cohomology.\footnote{In the case of non-abelian theories, one must appeal to a more general notion of cohomology \cite{etingof2005fusion,etingof2016tensor}.} Given a set of labels and fusion rules, a 3D TQFT with non-trivial labels/anyons is a cohomologically non-trivial solution to these polynomial equations.\footnote{In particular, we see that the space of consistent 3D TQFTs satisfying the MTC axioms is discrete \cite{etingof2016tensor}. In the context of Chern-Simons theory, this statement corresponds to the fact that there are no continuous deformations that preserve topological invariance and maintain a finite number of Wilson lines.} The quantities $(N_{ab}^c,R,F)$ are sometimes referred to as the \lq\lq MTC data," and $(S,T)$ are often referred to as the \lq\lq modular" data.

Crucially for us in what follows, the Pentagon and Hexagon equations are separable in a finite extension of $\mathds{Q}$. In other words, the solutions to these polynomials belong to a finite extension of the rational numbers, $K$ \cite{davidovich2013arithmetic}. This property enables us to use Galois theory to define a map, called Galois conjugation, from one TQFT to another.

\newsec{Galois conjugation}\label{galoisint}
Let us describe Galois conjugation in greater detail. To that end, denote the set of inequivalent modular solutions\footnote{One may in principle generalize this discussion to non-modular solutions, but our interest is in modular solutions.} of the Pentagon and Hexagon equations for a fixed set of fusion rules, $\rho$, as $\CS_{\rho}$. Since $(F,R)\in\CS_{\rho}$ explicitly involve fusion spaces, they depend on the corresponding bases. This phenomenon is called \lq\lq gauge dependence" in the MTC literature. As a result, the particular field extension, $K$, that $(F,R)$ take values in is gauge dependent. Given $K$, we can then define a gauge-dependent Galois group, ${\rm Gal}(K)$, for $(F,R)$ by demanding that it act as an automorphism of $K$ that fixes $\mathds{Q}$ pointwise.

As a simple example of a Galois action let us set aside TQFT for a moment and consider the simple polynomial equation, $x^2+1=0$. Extending the rationals by solutions of this equation yields $e=\mathds{Q}(i)$. The Galois group is isomorphic to $\mathds{Z}_2$, and this group acts via complex conjugation: $g(a+ib)=a-ib$ for non-trivial $g\in{\rm Gal}(\mathds{Q}(i))$. Here the Galois group takes one between the distinct solutions, $\pm i$, of $x^2+1=0$.

The Galois action on $(F,R)$ often acts in a similar fashion: it takes one between inequivalent modular solutions of the Pentagon and Hexagon equations (and hence inequivalent TQFTs) with the same fusion rules.\footnote{Therefore, if the theory has abelian anyons, the Galois action preserves the corresponding 1-form symmetry group.} On the other hand, since $(F,R)$ depend on gauge choices, one may also find a non-trivial Galois group that takes one between different gauge choices for the same theory.\footnote{For example, consider the so-called \lq\lq Toric Code" discrete gauge theory with $\mathds{Z}_2\times\mathds{Z}_2$ fusion rules. The four anyons can be labeled as $\left\{1,e,m,\epsilon\right\}$ and satisfy $e^2=m^2=\epsilon^2=1$, $em=\epsilon$ (with the rest of the non-trivial fusion rules following from this data). We may take $(F,R)$ to be
\begin{eqnarray}
R^1_{\epsilon,\epsilon}&=&R^{\epsilon}_{m,e}=R^e_{m,\epsilon}=R^m_{\epsilon,e}=-1~, \ \ \ R^{\epsilon}_{e,m}=R^1_{\epsilon,\epsilon}=R^1_{e,e}=R^1_{m,m}=1~,\cr F_{\epsilon,m,e}^1&=&F_{\epsilon,e,m}^1=-F_{m,e,\epsilon}^1=-F_{e,m,\epsilon}^1=i~,
\end{eqnarray}
with all other $F=1$. Clearly, the Galois group is $\mathds{Z}_2$ and acts via complex conjugation. However, this action is trivial in abelian group cohomology. Indeed, by rotating the basis vector $\psi\in V_{\epsilon,\epsilon}^1$ as  $\psi\to-i\psi$, we find that all $F,R=\pm1$. Therefore, in this gauge, the Galois group is trivial and so the original $\mathds{Z}_2$ Galois group acts trivially in group cohomology. Moreover, since all the quantum dimensions are unity, $\sqrt{\sum d_c}=2$, and so there is no non-trivial Galois action for this theory.
\label{toricCode}}

\tikzset{every picture/.style={line width=0.75pt}} 

\begin{figure}[h!]
\centering    
\begin{tikzpicture}[x=0.75pt,y=0.75pt,yscale=-0.8,xscale=0.8]

\draw    (175.25,178.78) .. controls (154.35,160.6) and (149.17,128.33) .. (188.31,110.89) ;

\draw    (147.82,193.33) .. controls (209.17,190.33) and (204.17,130.33) .. (180.25,117.76) ;

\draw    (147.82,193.33) .. controls (79.17,195.33) and (92.17,82.33) .. (171.17,112.33) ;

\draw    (188.31,110.89) .. controls (279.17,78.33) and (269.29,210.31) .. (185.69,187.27) ;

\draw    (453.25,179.78) .. controls (432.35,161.6) and (425.17,120.33) .. (466.31,111.89) ;

\draw    (425.82,194.33) .. controls (487.17,191.33) and (480.17,125.33) .. (458.25,118.76) ;

\draw    (425.82,194.33) .. controls (362.17,194.33) and (372.72,90.51) .. (447.17,112.33) ;

\draw    (466.31,111.89) .. controls (555.12,87.64) and (547.29,211.31) .. (463.69,188.27) ;

\draw    (490.17,108.33) .. controls (487.17,68.33) and (523.73,53.75) .. (540.45,98.04) .. controls (557.17,142.33) and (568.12,264.98) .. (425.82,194.33) ;

\draw (92,149) node   {$a$};
\draw (268,149) node   {$b$};
\draw (371,153) node   {$a$};
\draw (519,151) node   {$b$};
\draw (563,154) node   {$c$};
\draw (484,95) node   {$\mu $};
\draw (425,204) node   {$\nu $};
\draw (175,238) node   {$( a)$};
\draw (461,238) node   {$( b)$};

\end{tikzpicture}
\caption{(a) The $S$ matrix, and all the links we consider in this paper, are gauge invariant. (b) The punctured $S$-matrix ($\mu\ne0$) has self-intersections and is therefore gauge dependent.}\label{SSp}
\end{figure}
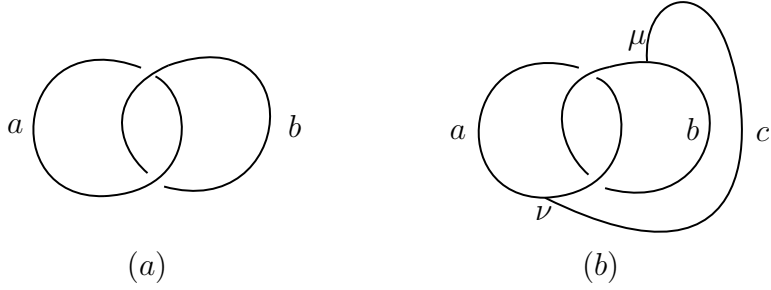

However, we can arrive at a more gauge-invariant notion of a Galois group as follows. Recall that a TQFT computes knot and link invariants. Given a link, $\CL$, its link invariant, $\CC(\mathcal{L})$, is written in terms of the MTC data. If $\CC(\CL)$ does not have self intersections, then it is independent of the choices of fusion space bases and hence is gauge invariant (see Fig. \ref{SSp}). Since we only consider observables built from such links in this paper, all our data will be gauge invariant. This fact implies that the link data we will study is defined over some gauge-invariant subfield, $\CK\subseteq K$. The corresponding Galois group, ${\rm Gal}(\CK)$, is then gauge independent.

As we will see, torus links will play a particularly important role in our story. These links can be constructed from words built out of the modular $S$ and $T$ matrices in \eqref{moddata}. For these matrices, the relevant field extension is a cyclotomic field, $\CK=\mathds{Q}(\xi_{N})$, given by extending the rationals by powers of a primitive root of unity, $\xi_{N}=\exp\left(2\pi i/ N\right)$ \cite{DeBoer:1990em,coste1994remarks,Coste:1999yc,bantay2003kernel,dong2015congruence}. As a result, the Galois group for the modular data is ${\rm Gal}(\mathds{Q}(\xi_{N}))=\mathds{Z}_{N}^{\times}$---the multiplicative group modulo $N$ consisting of all $n\in\left\{0,1,2,\cdots,N-1\right\}$ that are co-prime to $N$ (i.e., ${\rm gcd}(n,N)=1$).\footnote{Unfortunately, the $S$ and $T$ matrices, along with the topological central charge, are not enough to specify an MTC \cite{mignard2017modular}. As a result, we cannot take the Galois group of the $S$ and $T$ matrices to define a Galois group of the MTC in general. On the other hand, there may be other gauge-invariant ways to classify MTCs (e.g., see \cite{Bonderson_2019,wen2019distinguish} for preliminary results in this direction). Such a classification scheme might then allow one to assign a gauge-invariant Galois group for the full MTC.}
In this case, we have a trivial Galois action on $\mathds{Q}$ and a non-trivial action on $\xi$
\begin{equation}
q(\xi) = \xi^q~, \ \ \ \forall q\in\mathds{Z}_{N}^{\times}~.
\end{equation}
If the $S$ and $T$ matrices contain any elements not in $\mathds{Q}$, the Galois group will take the TQFT with modular data $(S, T)$ to a new TQFT with modular data $(q(S), q(T))$.\footnote{\label{gentrans}If $S$ or $T$ are not real, we can always take $(S,T)\to (S^*, T^*)$ (and similarly $(F,R)\to (F^*,R^*)$) and get a consistent TQFT related to the original one by time reversal.}

In fact, given this discussion, one can work out the Galois action on the modular data for a given $q\in\mathds{Z}_{N}^{\times}$\cite{coste1994remarks}\footnote{The Galois action on $T$ follows from the fact that $T_{ij}=\theta_{i}\delta_{ij}$ is a diagonal matrix of phases. The Galois action on $S$ follows from a careful analysis of the consequences of Verlinde's formula and the fact that the fusion rules are preserved by the Galois action.}
\be
\label{SandTGalois}
q(T_{aa})=(T_{aa})^q~, \ \ \     q(S_{ab})=\epsilon_{q}(a)S_{\sigma_q(a)b}=\epsilon_q(b)S_{a\sigma_q(b)}~,
\ee
where $\sigma(a)$ is a permutation of the labels and $\epsilon_{q}(a)\in \{\pm\}$. Hence, Galois conjugation of the $S$ matrix is a signed permutation.

We can say a bit more about how the Galois group acts on the modular data by making further contact with related results in the 2D RCFT literature \cite{bantay2003kernel}. In the context of RCFT, the natural normalization for the $T$ matrix is
\begin{equation}
T\to \varphi\cdot T~, \ \ \ \varphi=\exp(-\pi ic/12)~,
\end{equation}
where $c$ is the central charge, and $\varphi^3=\Theta$ (recall $\Theta$ was introduced in \eqref{sl2z}). In this normalization, Bantay showed that the Galois group is given as follows \cite{bantay2003kernel}
\begin{equation}\label{bantayres}
{\rm Gal}(\varphi\cdot T, S)={\rm Gal}(\mathds{Q}(\xi_N))\simeq\mathds{Z}_{N}^{\times}~,\ \ \ (\varphi\cdot T)^N=\mathds{1}~,
\end{equation}
where $N$ is the \lq\lq conductor"---for our purposes, the smallest $N>0$ such that $(\varphi\cdot T)^N$ is the identity matrix. By definition, Bantay's Galois group must have a (not necessarily faithful) action on $\varphi\cdot T_{00}=\varphi$. Therefore, going back to the natural MTC normalization for $T$, we may conclude that
\begin{equation}
{\rm Gal}(T, S)=\mathds{Z}_N^{\times}~,
\end{equation}
acts (not necessarily faithfully) on the modular data of the MTC and therefore constitutes a Galois group for the modular data. This statement does not preclude subgroups, $H\subset\mathds{Z}_N^{\times}$, from acting faithfully on the modular data of the MTC.\footnote{For example, in the context of the modular data of the Lee-Yang RCFT, the natural Galois group is $\mathds{Z}_{60}^{\times}\simeq\mathds{Z}_2^2\times\mathds{Z}_4$. This group acts unfaithfully on the modular data of the corresponding MTC. However, a $\mathds{Z}_{20}^{\times}\simeq\mathds{Z}_2\times\mathds{Z}_4\subset Z_{60}^{\times}$ subgroup does act faithfully on this data. Note that the $\mathds{Z}_{20}^{\times}$ subgroup is twice as large as the $\mathds{Z}_{5}^{\times}\simeq\mathds{Z}_4$ Galois group defined by the twists and quantum dimensions alone.}

In our discussion of entanglement entropy in non-abelian theories, we will see that a slightly different notion of the conductor arises. There it is more natural to discuss an \lq\lq MTC conductor" defined as the smallest $N_0>0$ such that
\begin{equation}\label{MTCcond}
T^{N_0}=\mathds{1}~.
\end{equation}
This quantity is closely related to Bantay's conductor since it turns out that $N=fN_0$, where $f\in\{$$1,$ $2$, $3$, $4$, $6$, $12\}$ \cite{bantay2003kernel}. 

While the above discussion, following \cite{bantay2003kernel}, is tied to the existence of RCFTs realizing a particular MTC, it turns out that one may rephrase the above discussion without explicit reference to an underlying RCFT \cite{dong2015congruence}. This latter approach is somewhat more mathematical, and we will not summarize it here. However, the upshot is that we will be able to make certain statements below about MTCs that need not be related to RCFTs.

Given our understanding of the Galois action on the modular data, let us briefly summarize the general picture of the Galois action on the full MTC. Note that a non-trivial Galois action on $S$ and $T$ is a sufficient but not necessary condition for the Galois action on the full MTC to be non-trivial. Indeed, the $F$ and $R$ matrices may still transform even if the modular data does not \cite{mignard2017modular}. We then have the following picture: the solutions of the Pentagon and Hexagon equations are partitioned into orbits of the Galois group.\footnote{Although these orbits generally have length greater than one, this is not always true. For example, the Galois orbit of the Toric Code MTC with $\mathds{Z}_2\times\mathds{Z}_2$ fusion rules is length one since the corresponding Galois group can always be taken to be trivial by an appropriate gauge choice (see the discussion in Footnote \ref{toricCode}).} 

\subsection{Explicit examples}\label{examples}
To illustrate the above discussion, let us consider two examples.

\bigskip
\noindent
{\bf Example 1: $\mathds{Z}_N$ TQFT ($N$ odd).} Let us consider abelian TQFTs with $\mathds{Z}_N$ fusion rules and $N$ odd.\foot{Note that these are not $\mathds{Z}_N$ discrete gauge theories. These latter theories have $\mathds{Z}_N\times\mathds{Z}_N$ fusion rules.} One set of solutions to the hexagon and pentagon equations is
\begin{equation}\label{ex1a}
F(j_1,j_2,j_3)=1~, \ \ \ R(j_1,j_2)=\exp\left({2\pi ij_1j_2\over N}\right)~,
\end{equation}
where $j_i\in\mathds{Z}_N$. From these quantities, we can build the modular data
\begin{equation}\label{ex1}
T(j_1,j_2)=\delta_{j_1,j_2}\exp\left({2\pi ij_1j_2\over N}\right)~,\ \ \ S(j_1,j_2)={1\over\sqrt{N}}R(j_1,j_2)R(j_2,j_1)={1\over\sqrt{N}}\exp\left({4\pi ij_1j_2\over N}\right)~.
\end{equation}
Clearly, for any $N$, the Galois action is non-trivial (i.e., the above solution always lies in a non-trivial Galois orbit). Moreover, it is straightforward to check that the modular data transforms as in \eqref{SandTGalois}.

As a particularly simple example, consider $N=3$. In this case, noting that $\sqrt{3}=\exp\left(2\pi i/12\right)+\exp\left(-2\pi i/12\right)$ makes clear that the Galois group is  $\mathds{Z}_{12}^{\times}\simeq\mathds{Z}_2\times\mathds{Z}_2$. Acting with $11\in\mathds{Z}_{12}^{\times}$ takes $(S,T)\to(S^*,T^*)$, while the remaining elements ($5,7\in\mathds{Z}_{12}^{\times}$) also flip the sign of the normalization of the $S$ matrix. The solution \eqref{ex1a}, \eqref{ex1}, with $N=3$ plugged in, corresponds to $SU(3)_1$ Chern-Simons theory. The Galois element $11\in\mathds{Z}_{12}^{\times}$ implements time reversal (as discussed in Footnote \ref{gentrans}) and produces $(E_6)_1$.\footnote{See the recent discussion in \cite{Cordova:2018qvg} where the time reversal relation between these theories was discussed.}

Note that $SU(3)_1$ and $(E_6)_1$ are unitary theories and correspond to abelian CS theories. In fact, as we will comment further below, it is more generally true that all abelian MTCs with $S$ as in \eqref{moddata} correspond to abelian CS theories.\footnote{On the other hand, if we flip the sign of the $S$-matrix, this statement is no longer true. This fact follows from a result of Milgram on Gauss sums (see \cite{milnor1973symmetric}).}

\bigskip
\noindent
{\bf Example 2: Fibonacci TQFT $\simeq(G_2)_1$ Chern-Simons.} Here we consider a non-abelian example. Let us suppose that there are two simple elements, $\{1,\tau\}$, and that the only non-trivial fusion rule is
\begin{equation}
\tau\otimes\tau=1+\tau~.
\end{equation}
The Fibonacci MTC, which gives rise to $(G_2)_1$ Chern-Simons theory, solves the pentagon and hexagon equations with these fusion rules. The corresponding non-trivial MTC data\footnote{All MTC data not explicitly mentioned is equal to 1.} is
\begin{eqnarray}
F_{\tau\tau\tau}^{\tau}&=&  \begin{pmatrix}
    \varphi^{-1} & \varphi^{-1/2}\\
    \varphi^{-1/2} & -\varphi^{-1}\\
  \end{pmatrix}~, \ \ \ R^1_{\tau\tau}=\xi^{2}~, \cr R^{\tau}_{\tau\tau}&=&\xi^{-{3\over2}}~, \ \varphi={1\over2}(1+\sqrt{5})=\xi^{-1}+1+\xi~, \ \xi=\exp\left({2\pi i\over5}\right)~.
\end{eqnarray}
From these quantities, one can construct the modular data
\begin{eqnarray}
S&=&{1\over\sqrt{2+\varphi}}  \begin{pmatrix}
    1 & \varphi\\
    \varphi & -1\\
  \end{pmatrix}~, \ \ \ T={\rm diag}\left(1,\exp\left({4\pi i\over5}\right)\right)~.
\end{eqnarray}
Writing $\sqrt{2+\varphi}=\exp\left(2\pi i/20\right)-\exp\left(2\pi i9/20\right)$, we see that a $\mathds{Z}_{20}^{\times}\subset\mathds{Z}_{60}^{\times}$ Galois group acts faithfully on the modular data (the $\mathds{Z}_{60}^{\times}$ group acts faithfully on the modular data in the RCFT normalization). The elements of $\mathds{Z}_{20}^{\times}$ are
\begin{eqnarray}
\mathds{Z}_{20}^{\times}\simeq\mathds{Z}_4\times\mathds{Z}_2=\left\{1,11\right\}\times\left\{1,3,7,9\right\}\simeq\left\{1,11\right\}\times\left\{1,7,49,43\right\}\subset\mathds{Z}_{60}^{\times}~.
\end{eqnarray}
Acting with $19\in\mathds{Z}_{20}^{\times}$ takes $\xi\to\xi^{19}=\bar\xi$ while leaving $S$ invariant and corresponds to time reversal. This transformation takes us to $(F_4)_1$. On the other hand, acting with $7,13\in\mathds{Z}_{4}\times\mathds{Z}_2$ flips the sign of the $S$-matrix and gives the Lee-Yang and conjugate Lee-Yang MTCs respectively (the remaining transformations give other theories related to the ones mentioned here by $S\to-S$).\footnote{See \cite{Dedushenko:2018bpp,Buican:2019huq} for examples of applications of the related $\mathds{Z}_5^{\times}\simeq\mathds{Z}_4$ Galois conjugation (acting on quantum dimensions and twists) to 4D superconformal field theories.}

\newsec{Multiboundary entanglement entropy in TQFT}\label{TEE}
In the previous section, we saw that solutions of the Pentagon and Hexagon equations can be partitioned into Galois orbits. This fact allows us to take the data of one TQFT and Galois conjugate it to get the data of another theory. The correlations functions / link invariants of the theory also get transformed under Galois conjugation. 

To make this abstract discussion somewhat more physical, we will study how the Galois action affects a particular type of entanglement entropy defined in \cite{balasubramanian2017multi,Balasubramanian2018EntanglementEA} (see also \cite{Salton:2016qpp}). As we will see, studying this question will lead to an interesting interplay between MTC data and the topology of 3-manifolds.

To proceed, let us imagine a unitary TQFT defined on a compact 3-manifold, $\CM_{\CL}$, that is a link complement of some closed 3-manifold, $\CM$. Note that we will mostly focus on the case $\CM=S^3$ in what follows (we briefly discuss certain generalizations to other Lens spaces in Sec. \ref{lens}). We can construct such an $\CM_{\CL}$ by first drawing a non-self-intersecting $n$-component link, $\CL^n=\sqcup_{i=1}^nL_i$, on $S^3$ and then removing a tubular neighborhood of the link, $\CN(\CL^n)$, from $S^3$. In other words
\begin{equation}
\CM_{\CL}\equiv S^3-\CN(\CL^n)~.
\end{equation}
Then, it is clear that $\partial\CM_{\CL}=\sqcup_{i=1}^nT^2_i$. In other words, the boundary of our 3-manifold consists of a disjoint union of $T^2$'s. 

To any $T^2$, we can associate a Hilbert space (we will discuss subtleties related to the case of non-unitary MTCs below), $\CH(T^2)$, whose basis states, $\{\ket{j_a}\}$, can be constructed by first filling in the $T^2$ to obtain a solid torus, $U$, with $\partial U= T^2$. The partition function of the theory on $U$ with line $j_a$ wrapping the non-contractible cycle of $U$,  $Z_U(j_a)$, then defines a corresponding state, $|j_a\rangle$, on the boundary $T^2$. We can compute the inner product of this state with a set of dual states by first thinking of $U$ as $U=D^2\times S^1$, where $D^2$ is the 2-disk. The dual state, $\langle j_b|$, comes from studying the partition function on $U'=D^2\times S^1$, where $\partial U'=-\partial U$,\footnote{In other words, $U$ and $U'$ share the same boundary with orientation reversed.} with a line, $j_b$, inserted along the non-contractible cycle. The corresponding inner product, $\langle j_b|j_a\rangle$, can also be obtained by instead inserting the conjugate line, $j_b^*$, (in addition to $j_a$) along the non-contractible cycle of $U$. The partition function for $U\cup U'=S^2\times S^1$ takes the form
\begin{equation}\label{innerprod}
\langle j_b|j_a\rangle\equiv Z_{S^2\times S^1}(j^*_b,j_a)=\delta_{ab}~,
\end{equation}
which follows from conservation of topological charge.

Now we can consider the boundary state, $\ket{\CL^n}$, which belongs to the tensor product of Hilbert spaces $\CH_1 \otimes \dots \otimes \CH_n$ corresponding to the $n$ $T^2$ boundaries. This state is defined by considering the partition function of the theory on $\CM_{\CL}$
\begin{equation}
|\CL^n\rangle\equiv Z_{\CM_{\CL}}~.
\end{equation}
We may expand this state in terms of the $T^2$ states as follows
\be\label{Lnstate}
\ket{\CL^n}= \sum_{j_1,...,j_n} C_{\CL^n}(j_1,...,j_n) \ket{j_1} \otimes\cdots\otimes \ket{j_n}~.
\ee
Using \eqref{innerprod}, we can compute the $C_{\CL^n}(j_1,\cdots,j_n)$ by considering the inner product with $\langle j_n|\otimes\cdots\otimes\langle j_1|$. As discussed above, this operation corresponds to filling in the boundary $T^2$'s and inserting conjugate representations, $j_i^*$, along the non-contractible cycles. For concreteness, let us consider a Chern-Simons theory and its Euclidean path integral on $\CM_{\CL}$. In this case, we have that the $\CC_{\CL^n}(j_1,\cdots,j_n)$ coefficients are just the various link invariants on $S^3$ computed from correlators of the Wilson lines in the conjugate representations
\begin{equation}\label{linkcoeffs}
C_{\CL^n}(j_1,\cdots,j_n) = \langle W_{j_1}^*\cdots W_{j_n}^*\rangle~.
\end{equation}

Given the link state, we can define the density matrix $\rho=\ket{\CL^n}\bra{\CL^n}$, where the coefficients of $\bra{\CL^n}$ are $\langle W_{j_1}\cdots W_{j_n}\rangle$. We can further define reduced density matrices of the form
\be
{\rho_{\rm red}}_{1,...,m}=\text{tr}_{m+1,...,n}(\rho)~,
\ee
where we trace over the Hilbert subspace $\mathcal{H}_{m+1} \otimes \cdots \otimes \mathcal{H}_n$ to get a matrix defined on $\mathcal{H}_1 \otimes \cdots \otimes \mathcal{H}_m$.  Then, one can define the MEE to be the usual von Neumann entropy of this reduced density matrix
\be\label{MEE}
S_{vN}(\rho_{\rm red})=- \text{tr}(\rho_{\rm red} \hspace{0.1cm} \text{ln} \rho_{\rm red})~.
\ee

This entanglement entropy is a coarse-grained form of the information contained in the link invariants $C_{\CL^n}(j_1,...,j_n)$.\footnote{Indeed, at a more operational level, one may simply view the MEE as a convenient and natural means to encapsulate information about the link invariants on $S^3$. This information can, in principle, be reconstructed without ever introducing boundaries and associated Hilbert spaces.} Many interesting properties of this entanglement entropy were studied in \cite{balasubramanian2017multi,Balasubramanian2018EntanglementEA,Salton:2016qpp} (see also \cite{Melnikov:2018zfn}). In the following section, we will compute the explicit form of the link state, $\ket{\CL^n}$, in general abelian theories and study the behavior of its entanglement entropy (after tracing out sub-links) under Galois conjugation. Note that since the entanglement entropy is invariant under local unitaries acting on the individual Hilbert spaces, we can ignore phases that come up in the calculation of $C_{\CL^n}(j_1,...,j_n)$ which depend purely on any one of the labels. Building on the results of the abelian discussion, we will then move on to discuss the more subtle case of non-Abelian TQFTs.

Before we continue, let us precisely define our procedure for comparing MEE under Galois transformations:

\bigskip\noindent
{\bf Definition 1 (comparing MEE under Galois conjugation):} By comparing the MEE under Galois transformations, what we mean is the following. We start with some unitary TQFT, $\CT$, and we compute the MEE. Then, we perform a Galois transformation to produce another TQFT, $\CT'$. We then compute the MEE in $\CT'$ and compare with the MEE in $\CT$. This comparison can be done directly by producing $\rho_{{\rm red}}(\CT')$ from $\rho_{{\rm red}}(\CT)$ via the Galois action. We then proceed iteratively along a Galois orbit, comparing MEEs for each element of the orbit. In particular, we do {\it not} apply a Galois transformation to \eqref{MEE} directly (this quantity is typically a transcendental number and does not lie in the field extension of the MTC).

\subsec{Subtleties for non-unitary theories}
In the next section, we will discuss abelian TQFTs. These theories are all described by (unitary) Abelian CS theories.\footnote{This statement ignores potential $S\to-S$ Galois transformations. However, we will see that our results apply to these theories as well.} In the language of axiomatic TQFT (e.g., see \cite{segal1988definition,walker1991witten,kontsevich1988rational}), they assign Hilbert spaces to boundaries of 3-manifolds, $\partial\CM$. In other words, to each boundary component of $\CM_{\CL}$, we have a complex vector space with a positive-definite norm.

On the other hand, when we discuss non-abelian TQFTs, the Galois action often takes unitary theories to non-unitary ones (as in Example 2 of Sec. \ref{examples}). Note that these non-unitary theories still have a finite number of simple objects. However, unlike unitary theories, non-unitary TQFTs have negative $S^3$ expectation values for some of the loops built out of the simple objects (e.g., a loop of the $\tau$ anyon in the Lee-Yang MTC discussed in Sec. \ref{examples} has $S^3$ expectation value $1-\varphi<0$). As a result, under the standard MTC Hermitian inner product, such theories have negative norm states.

Still, even for non-unitary theories, in the case of a 3-manifold with boundary 2-tori, $T^2_i$, the theory assigns vector spaces, $V(T^2_i)$, with a set of vectors obeying \eqref{innerprod}. Indeed, the existence of this pairing follows from topological charge conservation and is independent of unitarity. Moreover, the non-unitary theories we consider lie on the same Galois orbit as at least one unitary theory, so the link invariant coefficients in \eqref{linkcoeffs} and their orientation-reversed conjugates have a natural extension to the non-unitary case under the Galois action. As a result, even for the non-unitary theories we study, we may formally construct a positive semi-definite reduced density matrix as in the discussion below \eqref{Lnstate} for the state defined by the path integral over $\CM_{\CL}$.

Readers who find this discussion disturbing are encouraged to take the definition in the previous subsection as an operational definition for comparing MEE in our theories of interest. Note that for more general states it is not immediately clear to us if one can construct a reduced density matrix in the same way. However, in the context of related non-unitary 2D CFTs, like the Lee-Yang theory, it is known that one can construct standard density matrices for other closely related measures of entanglement and define a Hilbert space with respect to a modified norm \cite{Bianchini:2014uta}.\footnote{These ideas have also played a role in a non-unitary proof of Zamolodchikov's c-theorem \cite{Castro-Alvaredo:2017udm}.} We suspect that assigning such a Hilbert space to the subset of non-unitary MTCs we discuss here is also possible, but we do not prove it.\footnote{This statement may be related to the fact that the primaries in 2D CFTs like Lee-Yang have positive norm, while negative norms only enter at the level of the descendants. We thank A.~Konechny for discussions on this point.}

\newsec{Abelian TQFTs}\label{abeliansec}
In this section we will study how the multiboundary entanglement entropy described in the previous section transforms as we perform Galois conjugation on abelian TQFTs. As discussed in Sec. \ref{TQFTbasics}, abelian TQFTs have labels and fusion rules given by an abelian group, $\mathcal{A}$. Since the fusion rules are invariant under the Galois action, we see that the space of abelian TQFTs---and, more specifically, the space of theories with fusion rules given by $\CA$---is closed under Galois conjugation.

As we will discuss in more detail shortly, abelian TQFTs can always be written as abelian CS theories \cite{wall1963quadratic,wall1972quadratic,nikulin1980integral,Lee_2018}.\footnote{Here we ignore the possibility of flipping the sign of the $S$ matrix (as discussed below \eqref{sl2z}). However, our results apply even to any MTCs of this latter type.} Since the main topological property encoded by such theories is linking number, it is intuitively reasonable to imagine that the Galois action will lead to abelian theories with the same entanglement entropy.\footnote{In this case, the simpler entanglement entropy of \cite{kitaev2006topological,Levin:2006zz} is trivially invariant since it is given by the square-root of the rank of the fusion group, $\sqrt{|\CA|}$.} We will indeed see this expectation is correct and that the linear transformation properties of $S$ under the Galois action \eqref{SandTGalois} play an important role.

To proceed, let us first discuss abelian TQFTs in more detail. Since the fusion rules are those of an abelian group, the fusion coefficients satisfy $N_{ab}^c=\delta_{a\cdot b, c}$ where $a,b,c \in \mathcal{A}$, and $a\cdot b$ is the group multiplication. Moreover, $N_{ab}^c \in \{0, 1\}$, and so all fusion spaces are one dimensional. Hence, the $F$ and $R$ matrices are just phases, and we will denote them as $F(a,b,c)$ and $R(a,b)$.\footnote{Note that since all fusion processes are one dimensional, specifying $a,b,c$ in $F_{a,b,c}^d$ automatically specifies $d$, and so we loose no generality in taking the $F$ symbols to depend on three group elements. Similar reasoning shows that we loose no generality in taking the $R$ matrices to depend on two group elements.}

In this case, the pentagon equation simplifies to
\be
\label{AbelianPentagon}
F(a,b,c\cdot d)F(a\cdot b,c,d)=F(b,c,d)F(a,b\cdot c,d)F(a,b,c)~.
\ee
A function $F:\CA \otimes \CA \otimes \CA \rightarrow U(1)$  satisfying \eqref{AbelianPentagon} is called a 3-cocyle in group cohomology. Similarly, the Hexagon equations reduce to 
\bea
\label{AbelianHexagon}
R(a,c)F(b,a,c)R(a,b)&=&F(b,c,a)R(a,b \cdot c)F(a,b,c)~, \nonumber\\
R(c,a)F(b,a,c)^{-1}R(b,a)&=&F(b,c,a)^{-1}R(b \cdot c,a)F(a,b,c)^{-1}~.
\eea 
The gauge-inequivalent solutions, $(F,R)$, belong to the third abelian cohomology group, $H^3_{ab}(\CA,U(1))$. The gauge freedom in $F$ and $R$ is captured by this cohomology structure \cite{etingof2005fusion,etingof2016tensor}. 

As reviewed in Sec. \ref{TQFTbasics}, to find the MTC data of a general TQFT given a set of labels and fusion rules, one finds $F$ matrices solving the Pentagon equations and then one solves the Hexagon equations given these $F$ matrices. However, in abelian TQFTs, the situation is much simpler, and the MTC data is fixed by the choice of a quadratic function, $\theta(a): \CA \rightarrow U(1)$, that gives the topological spins (i.e., the $T$ matrix).\footnote{For further details, see the recent discussion in \cite{Lee_2018}.}

Although much of what we said above does not depend on the existence of Lagrangians, it will be useful for us to keep them in mind in our subsequent discussion of abelian theories. Moreover, as mentioned at the beginning of this section, it turns out that we do not lose any generality in studying abelian Chern-Simons theories with gauge group $U(1)^N$ \cite{wall1963quadratic,wall1972quadratic,nikulin1980integral,Lee_2018} (they span the space of TQFTs with $S$ matrices as in \eqref{moddata}). These theories have Lagrangians of the general form
\be
S=\frac{iK_{ij}}{2\pi}\int_M A^{i}dA^{j}~.
\ee
Here $A^{i}$ are $U(1)$ gauge fields, $M$ is a 3-manifold, and $K$ is a symmetric even integral matrix of levels.\footnote{In other words, we will assume that the diagonal entries in $K$ are even integers (the remaining entries may be even or odd). Otherwise, the theory would be a spin-TQFT.} The fusion rules of the theory are given by the abelian group, $\mathds{Z}^N/K\mathds{Z}^N$, and anyons are labelled by a  set of basis vectors for this lattice. The fact that $K$ is an integer matrix means it has a Smith normal form, $K_S$, which we denote by 
\be
K_S=\begin{pmatrix}
n_1 & 0 & \dots & 0 \\
0 &n_2 &  \\
\vdots & & \ddots   & \\
0 & \dots  &0  &n_{N}
\end{pmatrix}~.
\ee
From this discussion, it is clear that the abelian group $\mathds{Z}^N/K\mathds{Z}^N$ is isomoprphic to $\mathds{Z}_{n_1} \otimes \cdots \otimes \mathds{Z}_{n_N}$. Clearly we can reproduce any finite abelian fusion group using such theories. As discussed above, the MTC data is specified by the topological spin. For abelian theories, it can be expressed in terms of $K$ as follows
\be
\theta(\vec{a})= \exp\left({\pi i \vec{a} K^{-1} \vec{a}}\right)~,
\ee
where $\vec{a}\in\mathds{Z}^N/K\mathds{Z}^N$.

Next let us explicitly fix the remainder of the modular data (recall that $T$ is given in terms of $\theta$). To that end, we first define the braiding 
\be\label{Bab}
B(\vec{a}.\vec{b})=\frac{\theta(\vec{a}+\vec{b})}{\theta(\vec{a})\theta(\vec{b})}  =\exp\left({2\pi i \vec{a} K^{-1} \vec{b}}\right)~.
\ee
The $R$ matrices and the braiding phase are related by \cite{Lee_2018}
\be
B(\vec{a},\vec{b})=R(\vec{a},\vec{b})R(\vec{b},\vec{a})~.
\ee
The representation of the modular group generators $S$ and $T$ realized by this theory is then
\be
\label{SandTfromTheta}
S_{\vec{a},\vec{b}}=\frac{1}{\sqrt{|\CA|}} B(\vec{a},\vec{b})~, \ \ \ T_{\vec{a},\vec{a}}=\theta(\vec{a})~,
\ee
where $|\CA|$ is the order of the abelian group, $\CA$. 

In the next section, we will use the above data to find the link invariant for a general $n$-component link. Given this expression, we will then compute the entanglement entropy for general abelian theories and show the invariance claimed above under Galois transformations.

\subsection{Link invariants in abelian TQFTs}
Let us consider an $n$-component link in which the constituent knots are labelled by $j_1,\cdots,j_n$. Since the $F$ and $R$ matrices of abelian TQFTs are $U(1)$ valued, simplifying the individual structure of a knot, $j_i$, to give the unknot will give us phases which act on the Hilbert space, $\CH_i$. Since these phases can be removed using a local unitary operation, the entanglement entropy is independent of these phases. Hence, as far as calculating the entanglement entropy is concerned, we are only interested in the braiding between the constituent knots. We will consider the case of a 2-component link in the next section which can then be easily generalized to give the link invariant for an $n$-component link. 

\subsubsection{Link invariant for a 2-link}
\label{sec2-link}

Since abelian theories primarily capture linking number, it is reasonable to imagine that any link invariant can be written (up to unimportant local unitary transformations that will not affect our quantities of interest) in terms of the $S$ matrix. We will see this statement is indeed true.

\tikzset{every picture/.style={line width=0.75pt}} 
\begin{figure}[h!]
\centering
\begin{tikzpicture}[x=0.75pt,y=0.75pt,yscale=-1,xscale=1]

\draw    (155.17,61.33) .. controls (119.17,78.33) and (111.17,112.33) .. (133.17,128.33) ;

\draw    (137.17,76.33) .. controls (164.17,91.33) and (148.17,130.33) .. (120.17,140.33) ;

\draw    (116.17,60.33) -- (132.17,72.33) ;

\draw    (138.17,132.33) -- (151.17,141.33) ;

\draw   (124.29,62.8) -- (125.29,66.76) -- (121.62,68.55) ;
\draw   (146.1,70.4) -- (142.12,69.51) -- (142.16,65.43) ;
\draw   (132.11,137.89) -- (128.12,137.02) -- (128.14,132.94) ;
\draw   (145.61,133.83) -- (146.23,137.86) -- (142.42,139.31) ;
\draw    (259.17,61.33) .. controls (223.17,78.33) and (215.17,112.33) .. (237.17,128.33) ;

\draw    (241.17,76.33) .. controls (268.17,91.33) and (252.17,130.33) .. (224.17,140.33) ;

\draw    (220.17,60.33) -- (236.17,72.33) ;

\draw    (242.17,132.33) -- (255.17,141.33) ;

\draw   (228.29,62.8) -- (229.29,66.76) -- (225.62,68.55) ;
\draw   (244.72,66.28) -- (248.31,68.22) -- (247.15,72.14) ;
\draw   (236.11,137.89) -- (232.12,137.02) -- (232.14,132.94) ;
\draw   (248.55,140.59) -- (248.03,136.54) -- (251.88,135.19) ;
\draw    (359.17,62.33) .. controls (323.17,79.33) and (315.17,113.33) .. (337.17,129.33) ;

\draw    (341.17,77.33) .. controls (368.17,92.33) and (352.17,131.33) .. (324.17,141.33) ;

\draw    (320.17,61.33) -- (336.17,73.33) ;

\draw    (342.17,133.33) -- (355.17,142.33) ;

\draw   (326.53,70.24) -- (327.29,66.23) -- (331.37,66.15) ;
\draw   (350.1,71.4) -- (346.12,70.51) -- (346.16,66.43) ;
\draw   (330.24,135.37) -- (334.17,136.48) -- (333.9,140.55) ;
\draw   (349.61,134.83) -- (350.23,138.86) -- (346.42,140.31) ;
\draw    (462.17,61.33) .. controls (426.17,78.33) and (418.17,112.33) .. (440.17,128.33) ;

\draw    (444.17,76.33) .. controls (471.17,91.33) and (455.17,130.33) .. (427.17,140.33) ;

\draw    (423.17,60.33) -- (439.17,72.33) ;

\draw    (445.17,132.33) -- (458.17,141.33) ;

\draw   (429.52,69.24) -- (430.3,65.23) -- (434.38,65.16) ;
\draw   (446.9,67.56) -- (450.98,67.76) -- (451.63,71.79) ;
\draw   (433.08,134.65) -- (437.1,135.38) -- (437.22,139.46) ;
\draw   (451.89,140.63) -- (450.97,136.66) -- (454.66,134.92) ;

\draw (115,50) node   {$a$};
\draw (117.17,147.33) node   {$a$};
\draw (159,53) node   {$b$};
\draw (156,148) node   {$b$};
\draw (219,50) node   {$a$};
\draw (221.17,147.33) node   {$a$};
\draw (263,53) node   {$b$};
\draw (260,148) node   {$b$};
\draw (319,51) node   {$a$};
\draw (321.17,148.33) node   {$a$};
\draw (363,54) node   {$b$};
\draw (360,149) node   {$b$};
\draw (422,50) node   {$a$};
\draw (424.17,147.33) node   {$a$};
\draw (466,53) node   {$b$};
\draw (463,148) node   {$b$};
\draw (135,175) node   {$( i)$};
\draw (238,176) node   {$( ii)$};
\draw (338,175) node   {$( iii)$};
\draw (442,176) node   {$( iv)$};
\end{tikzpicture}
\caption{Possible braidings for oriented links.}
\label{possbr}
\end{figure}
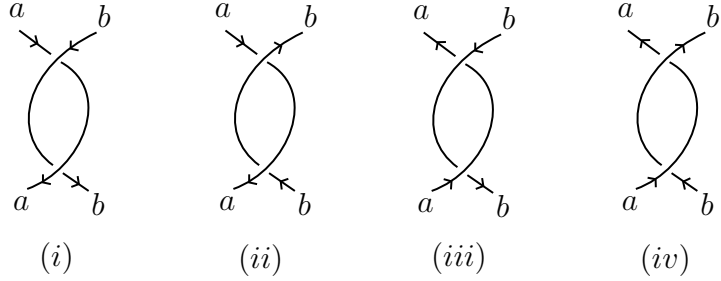
To that end, consider a 2-component link in which the two knots are labelled $a$ and $b$. There must be an even number of braids between them. As a result, the braids can be grouped into pairs. In an oriented link, four types of pairs are possible (see Fig. \ref{possbr}). Let us find the algebraic expression obtained from unbraiding diagram $(i)$ in Fig. \ref{possbr}.

\tikzset{every picture/.style={line width=0.75pt}} 

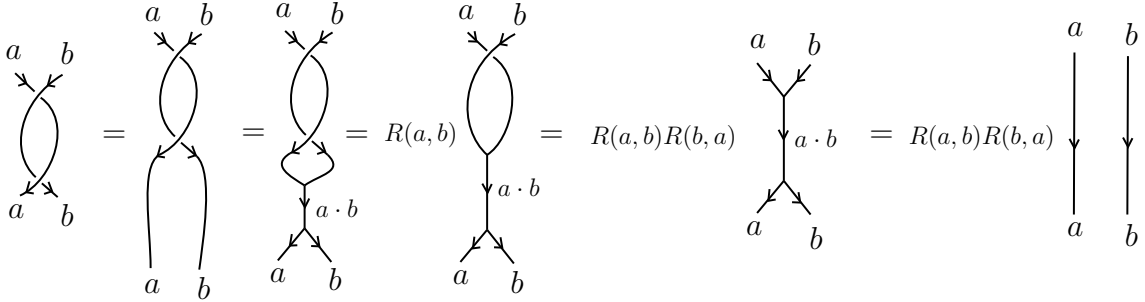
\begin{figure}[h!]
    \centering
\begin{tikzpicture}[x=0.75pt,y=0.75pt,yscale=-1.3,xscale=0.9]

\draw    (46.96,67.9) .. controls (22.42,78) and (16.97,98.2) .. (31.96,107.71) ;

\draw    (34.69,76.81) .. controls (53.09,85.72) and (42.19,108.89) .. (23.1,114.83) ;

\draw    (20.38,67.31) -- (31.28,74.44) ;

\draw    (35.37,110.08) -- (44.23,115.43) ;

\draw   (24.59,67.5) -- (26.59,71.16) -- (22.1,72.19) ;
\draw    (124.93,52.09) .. controls (100.4,62.19) and (94.95,82.39) .. (109.94,91.89) ;

\draw    (112.67,61) .. controls (131.07,69.91) and (120.16,93.08) .. (101.08,99.02) ;

\draw    (98.35,51.49) -- (109.26,58.62) ;

\draw    (113.35,94.27) -- (122.21,99.62) ;

\draw    (96.31,142.99) .. controls (96.31,127.54) and (90.17,103.78) .. (101.08,99.02) ;

\draw    (123.57,143.58) .. controls (125.62,132.89) and (131.75,106.15) .. (122.21,99.62) ;

\draw    (198.27,51.49) .. controls (173.74,61.59) and (168.28,81.79) .. (183.28,91.3) ;

\draw    (186,60.41) .. controls (204.41,69.32) and (193.5,92.49) .. (174.42,98.43) ;

\draw    (171.69,50.9) -- (182.6,58.03) ;

\draw    (186.69,93.68) -- (195.55,99.02) ;

\draw    (182.6,110.9) .. controls (174.42,107.34) and (163.51,103.18) .. (174.42,98.43) ;

\draw    (182.6,110.9) .. controls (190.78,106.75) and (205.09,105.56) .. (195.55,99.02) ;

\draw    (182.6,110.9) -- (182.6,127.54) ;

\draw    (182.6,127.54) -- (167.6,140.02) ;

\draw    (182.6,127.54) -- (197.59,140.61) ;

\draw    (300.56,52.09) .. controls (267.84,64.56) and (268.52,88.92) .. (284.88,99.02) ;

\draw    (288.29,61) .. controls (306.69,69.91) and (298.51,93.08) .. (284.88,99.02) ;

\draw    (273.97,51.49) -- (284.88,58.62) ;

\draw    (284.88,99.02) -- (284.88,128.13) ;

\draw    (284.88,128.13) -- (269.89,140.61) ;

\draw    (284.88,128.13) -- (299.87,141.2) ;

\draw    (451.16,75.85) -- (451.16,109.12) ;

\draw    (451.16,109.12) -- (436.17,121.6) ;

\draw    (451.16,109.12) -- (466.15,122.19) ;

\draw    (436.17,63.38) -- (451.16,76.45) ;

\draw    (466.15,63.97) -- (451.16,76.45) ;

\draw    (614.01,59.28) -- (613.69,122.19) ;

\draw    (643.99,60.69) -- (643.65,122.19) ;

\draw   (42.83,72.52) -- (38.69,72.05) -- (40.44,67.78) ;
\draw   (39.09,109.5) -- (41.09,113.16) -- (36.6,114.19) ;
\draw   (30.83,114.02) -- (26.69,113.55) -- (28.44,109.28) ;
\draw   (102.59,51.5) -- (104.59,55.16) -- (100.1,56.19) ;
\draw   (119.59,95.5) -- (121.59,99.16) -- (117.1,100.19) ;
\draw   (176.09,51) -- (178.09,54.66) -- (173.6,55.69) ;
\draw   (189.86,130.51) -- (190.87,134.55) -- (186.26,134.41) ;
\draw   (277.86,51.01) -- (278.87,55.05) -- (274.26,54.91) ;
\draw   (292.36,131.01) -- (293.37,135.05) -- (288.76,134.91) ;
\draw   (192.09,94) -- (194.09,97.66) -- (189.6,98.69) ;
\draw   (121.83,56.02) -- (117.69,55.55) -- (119.44,51.28) ;
\draw   (104.33,100.02) -- (100.19,99.55) -- (101.94,95.28) ;
\draw   (179.83,98.52) -- (175.69,98.05) -- (177.44,93.78) ;
\draw   (185.21,115.97) -- (182.77,119.35) -- (179.9,115.74) ;
\draw   (195.33,55.52) -- (191.19,55.05) -- (192.94,50.78) ;
\draw   (296.33,56.52) -- (292.19,56.05) -- (293.94,51.78) ;
\draw   (281.07,134.94) -- (276.92,135.33) -- (277.75,130.8) ;
\draw   (287.48,110.33) -- (285.1,113.75) -- (282.17,110.19) ;
\draw   (442.86,66.01) -- (443.87,70.05) -- (439.26,69.91) ;
\draw   (459.36,113.51) -- (460.37,117.55) -- (455.76,117.41) ;
\draw   (463.08,69.9) -- (458.94,70.35) -- (459.7,65.81) ;
\draw   (446.59,116.35) -- (442.46,116.87) -- (443.15,112.31) ;
\draw   (453.52,89.55) -- (451.06,92.92) -- (448.22,89.29) ;
\draw   (616.52,93.05) -- (614.06,96.42) -- (611.22,92.79) ;
\draw   (646.52,92.55) -- (644.06,95.92) -- (641.22,92.29) ;
\draw   (178.51,134.12) -- (174.35,134.25) -- (175.45,129.78) ;

\draw (19.58,59.17) node   {$a$};
\draw (21.06,120.99) node   {$a$};
\draw (49.57,59.31) node   {$b$};
\draw (49.53,122.39) node   {$b$};
\draw (97.56,44.36) node   {$a$};
\draw (127.55,44.5) node   {$b$};
\draw (126.14,150.73) node   {$b$};
\draw (170.9,42.94) node   {$a$};
\draw (200.89,43.9) node   {$b$};
\draw (273.18,43.54) node   {$a$};
\draw (303.17,44.5) node   {$b$};
\draw (436.73,54.83) node   {$a$};
\draw (468.09,56.02) node   {$b$};
\draw (97.31,149.96) node   {$a$};
\draw (167.6,145.81) node   {$a$};
\draw (199.72,147.4) node   {$b$};
\draw (247.74,91.2) node [scale=0.8]  {$R( a,b)$};
\draw (383.63,92.47) node  [scale=0.8]  {$R( a,b) R( b,a)$};
\draw (614.57,50.1) node   {$a$};
\draw (645.92,52.1) node   {$b$};
\draw (562.13,92.29) node  [scale=0.8]  {$R( a,b) R( b,a)$};
\draw (270.2,146.33) node   {$a$};
\draw (434.65,127.3) node   {$a$};
\draw (614.57,127.93) node   {$a$};
\draw (301.7,148.22) node   {$b$};
\draw (469.28,131.19) node   {$b$};
\draw (646,130.19) node   {$b$};
\draw (200.55,119.67) node [scale=0.8]   {$a\cdot b$};
\draw (302.55,111.67) node [scale=0.8]   {$a\cdot b$};
\draw (468.89,92.54) node [scale=0.8]   {$a\cdot b$};
\draw (506,92) node   {$=$};
\draw (321,92) node   {$=$};
\draw (212,92) node   {$=$};
\draw (154,91) node   {$=$};
\draw (76,92) node   {$=$};

\end{tikzpicture}
\caption{Link invariant of a braid pair. Diagrams 1-6 from left.}
\label{bpairtrans}   
\end{figure}

To understand this unbraiding, consider Fig. \ref{bpairtrans}. In going from diagram 2 to 3 (from left) of Fig. \ref{bpairtrans}, we have used the decomposition of the identity, which has a unique channel in an abelian TQFT. In diagrams 4 and 5, we remove the braiding by adding appropriate $R$ matrix factors. Finally we again use the decomposition of the identity to go from diagram 5 to 6. As a result, we find that the braid pair can be replaced by the identity acting on the two anyons if we include the factor $R(a,b)R(b,a)$.
\tikzset{every picture/.style={line width=0.75pt}} 

\begin{figure}[h!]
\centering
\begin{tikzpicture}[x=0.75pt,y=0.75pt,yscale=-0.95,xscale=0.95]

\draw    (188.25,176.78) .. controls (167.35,158.6) and (162.17,126.33) .. (201.31,108.89) ;

\draw    (160.82,191.33) .. controls (222.17,188.33) and (217.17,128.33) .. (193.25,115.76) ;

\draw    (160.82,191.33) .. controls (92.17,193.33) and (105.17,80.33) .. (184.17,110.33) ;

\draw    (201.31,108.89) .. controls (292.17,76.33) and (282.29,208.31) .. (198.69,185.27) ;

\draw   (165.58,101.1) .. controls (167.01,103.48) and (168.54,104.98) .. (170.15,105.58) .. controls (168.45,105.86) and (166.66,107.02) .. (164.78,109.06) ;
\draw   (206.91,110.39) .. controls (204.32,109.39) and (202.2,109.15) .. (200.55,109.66) .. controls (201.72,108.4) and (202.43,106.38) .. (202.65,103.62) ;
\draw   (174.33,193.3) .. controls (172.04,191.74) and (170.02,191.02) .. (168.3,191.15) .. controls (169.73,190.18) and (170.88,188.38) .. (171.73,185.74) ;
\draw   (212.76,183.02) .. controls (214.09,185.46) and (215.55,187.01) .. (217.14,187.69) .. controls (215.43,187.89) and (213.58,188.97) .. (211.62,190.93) ;
\draw    (427.69,185.27) .. controls (403.17,177.33) and (392.17,136.33) .. (413.31,118.89) ;

\draw    (409.17,180.33) .. controls (306.17,224.33) and (334.17,64.33) .. (413.17,110.33) ;

\draw    (424.17,111.33) .. controls (515.03,78.78) and (511.29,208.31) .. (427.69,185.27) ;

\draw   (397.27,107.77) .. controls (396.11,105.24) and (394.76,103.59) .. (393.22,102.81) .. controls (394.95,102.72) and (396.86,101.77) .. (398.95,99.95) ;
\draw   (441.39,110.22) .. controls (439.06,108.71) and (437.04,108.04) .. (435.32,108.2) .. controls (436.73,107.2) and (437.84,105.38) .. (438.63,102.72) ;
\draw   (387.82,183.6) .. controls (390.21,185.02) and (392.27,185.6) .. (393.97,185.37) .. controls (392.61,186.42) and (391.58,188.3) .. (390.9,190.99) ;
\draw   (441.76,183.02) .. controls (443.09,185.46) and (444.55,187.01) .. (446.14,187.69) .. controls (444.43,187.89) and (442.58,188.97) .. (440.62,190.93) ;
\draw    (413.17,110.33) .. controls (443.17,125.33) and (425.17,167.33) .. (417.17,172.33) ;

\draw (105,147) node    {$\vec j_1$};
\draw (278,147) node    {$\vec j_2$};
\draw (52,148) node    {$B(\vec j_1,\vec j_2) \ =\ $};
\draw (331,147) node    {$\vec j_1$};
\draw (505,147) node    {$\vec j_2$};
\draw (308,148) node    {$=\ $};
\draw (573,148) node    {$=B\left( \vec j_1^{*} ,\vec j_2\right)^{-1} \ \ $};
\end{tikzpicture}
\caption{The relation in \eqref{invrel} follows from the equality of the above TQFT diagrams.}\label{invdiagram}
\end{figure}

In the $K$ matrix formalism, the knots are labelled by anyonic vectors $\vec{j_1},\vec{j_2}\in \mathds{Z}^N/K\mathds{Z}^{N}$, and we think of the anyons as elements of the corresponding additive group. In this notation, the algebraic expression corresponding to the diagram $(i)$ is $R(\vec{j_1},\vec{j_2})R(\vec{j_2},\vec{j_1})$. From \eqref{SandTfromTheta}, we know that this is just the braiding phase $B(\vec{j_1},\vec{j_2})$. Repeating the above calculation for diagrams $(ii)$, $(iii)$ and $(iv)$ we get  $B(\vec{j_1},\vec{j_2^*})$, $B(\vec{j_1^*},\vec{j_2})$, and $B(\vec{j_1^*},\vec{j_2^*})$, respectively. If there are $n_1$ braid pairs of type $(i)$, $n_2$ of type $(ii)$, $n_3$ of type $(iii)$, and $n_4$ of type $(iv)$, the total link invariant is given by  
\be
(B(\vec{j_1},\vec{j_2}))^{n_1}  (B(\vec{j_1},\vec{j_2^*}))^{n_2} (B(\vec{j_1^*},\vec{j_2}))^{n_3} (B(\vec{j_1^*},\vec{j_2^*}))^{n_4} ~.
\ee
Moreover, \eqref{Bab} implies the following relations hold
\be\label{invrel}
B(\vec{j_1},\vec{j_2^*})=(B(\vec{j_1},\vec{j_2}))^{-1}~, \ \ \  B(\vec{j_1^*},\vec{j_2})=(B(\vec{j_1},\vec{j_2}))^{-1}~,
\ee
since $\vec{j_i^*}\sim-\vec{j_i}$, where \lq\lq$\sim$" means, \lq\lq up to vectors of the form $K\cdot\vec{\omega_i}$" (i.e., up to a $K$-trivial vector). In fact, \eqref{invrel} holds without the need to appeal to a $K$ matrix, since the TQFT diagrams in Fig. \ref{invdiagram} are equal.

Hence, the link invariant simplifies to
\be
\label{2linkinvariant}
B(\vec{j_1},\vec{j_2})^{l_{12}}\sim S(\vec{j_1},\vec{j_2})^{l_{12}}~,
\ee
where $l_{12}=n_1+n_4-n_2-n_3$ is the linking number, and \lq\lq$\sim$" means, \lq\lq up to an overall normalization." This simple calculation shows that the MTC data of abelian TQFTs can be used to compute the linking number of a link and that the result can be expressed through the $S$ matrix alone. Next, we generalize this argument to an $n$-link.

\subsubsection{Link invariant for an n-link}
For a link made up of $n$ knots, we should repeat the calculation in Sec. \ref{sec2-link} for each pair of knots, $(j_i,j_k)$, where $1\leq i<k\leq n$, and $j_{i,k}$ are the labels of the corresponding knots. Proceeding in this way, we find
\be
(B(\vec{j}_i, \vec{j}_k))^{\ell_{ik}}~,
\ee
where $\ell_{ik}$ is the linking number between the knots labelled $j_i$ and $j_k$ in the link. The total link invariant will be the product of these factors. As a result, the link invariant for an $n$-link in an abelian TQFT is
\be
\label{nlinkinvariant}
\prod_{i<k} (B(\vec{j}_i, \vec{j}_k))^{\ell_{ik}}~.
\ee

\subsection{Galois conjugation of entanglement entropy}
Using the link invariants computed in the previous subsection, we can now find the associated entanglement entropy defined in Sec. \ref{TEE} and study its behavior under the Galois action.

Let us again specialize to a 2-link before discussing the general $n>2$ case. To that end, using \eqref{2linkinvariant}, we have the normalized link state
\be\label{2linkstate}
\ket{\mathcal{L}^2}= \frac{1}{|\CA|} \sum_{\vec{j_1},\vec{j_2}} (B(\vec{j_1},\vec{j_2}))^{l_{12}} \ket{j_1,j_2} = \sum_{\vec{j_1},\vec{j_2}} |\CA |^{\frac{l_{12}}{2}-1} \Big (S_{\vec{j_1},\vec{j_2}}\Big  )^{l_{12}} \ket{j_1,j_2}~.
\ee
Tracing out the Hilbert space of the second link yields the following reduced density matrix
\bea
\rho_{red}&=&\sum_{\vec{j_1},\vec{h_1}} \sum_{\vec{m}} |\CA |^{l_{12}-2} \Big (S_{\vec{j}_1 \vec{m}} \Big )^{l_{12}} \Big (S_{\vec{h}_1 \vec{m}} \Big )^{-l_{12}} \ket{\vec{j_1}} \bra{\vec{h_1}} ~.
\eea 
Next we may use \eqref{SandTGalois} to perform a Galois transformation and read off the reduced density matrix in the conjugated theory 
\bea
\rho_{red}&=& |\CA |^{l_{12}-2} \sum_{\vec{j_1},\vec{h_1}} \sum_{\vec{m}} \Big ( \epsilon_p(m) S_{\vec{j}_1 \sigma_p(\vec{m})} \Big )^{l_{12}} \Big ( \epsilon_p(m) S_{\vec{h}_1 \sigma_p(\vec{m})} \Big )^{-l_{12}} \ket{\vec{j_1}} \bra{\vec{h_1}} \nonumber \\
&=& |\CA |^{l_{12}-2} \sum_{\vec{j_1},\vec{h_1}} \sum_{\vec{m}}  \Big (S_{\vec{j}_1 \sigma_p(\vec{m})} \Big )^{l_{12}} \Big (S_{\vec{h}_1 \sigma_p(\vec{m})} \Big )^{-l_{12}} \ket{\vec{j_1}} \bra{\vec{h_1}} ~.
\eea
Since $\vec{m}$ is summed over, the reduced density matrix is invariant under Galois conjugation. As a result, the entanglement entropy for a 2-link computed in an abelian TQFT and the Galois conjugated theory are equal.

The generalization to an $n$-link is straightforward. Indeed, using \eqref{nlinkinvariant}, the link state is given (up to a normalization factor) by
\bea\label{Lnstate}
\ket{\mathcal{L}^n}&=&  \sum_{\vec{j_1},...,\vec{j_n}} \prod_{j_i\leq j_k} \Big (S_{\vec{j_i},\vec{j_k}}\Big  )^{\ell_{ik}} \ket{j_1,...,j_n}~.
\eea
The density matrix for this state is then
\bea
\rho= \sum_{\vec{j_1},...,\vec{j_n}} \sum_{\vec{h_1},...,\vec{h_n}} \bigg(\prod_{i<k} \Big (S_{\vec{j}_i \vec{j}_k} \Big )^{\ell_{ik}}\bigg ) \bigg(\prod_{z<w} \Big (S_{\vec{h}_z \vec{h}_w} \Big )^{-\ell_{zw}}\bigg) \ket{\vec{j_1},...,\vec{j_n}} \bra{\vec{h_1},...,\vec{h_n}}~.
\eea
Without loss of generality, we can trace over the last $n-q$ links to get a reduced density matrix over the first $q$ links. Up to overall constant factors and phases which can be removed by applying unitaries on the first $q$ copies of the Hilbert space (which again don't affect the entanglement entropy), the components of the reduced density matrix can be written as  
\bea
\label{qlredden}
{\rho_{\rm red}}_{\vec{j_1}...\vec{j_q},\vec{h_1},...,\vec{h_q}}&=& \prod_{k=q+1}^n \sum_{\vec{m}} \prod_{i=1}^q \Big (S_{\vec{j_i},\vec{m}} \Big )^{\ell_{ik}} \Big (S_{\vec{h_i},\vec{m}} \Big )^{-\ell_{ik}} ~.
\eea
Galois conjugation of this reduced density matrix will only result in a permutation of the vectors $\vec{m}$. Since there is a sum over $\vec{m}$, the reduced density matrix is invariant under Galois conjugation. Note that this invariance includes any Galois transformation taking $S\to-S$ (as in Example 1 of Sec. \ref{examples}).\footnote{In fact, we could have constructed a link state directly for such theories using diagramatic reductions analogous to those above, and we would have found the same reduced density matrix. Therefore, such theories have invariant $\rho_{\rm red}$ even if they are not Galois conjugates of abelian CS theories.}

This discussion once again implies that the entanglement entropy is also invariant under Galois conjugation. Thus, even though the link invariants calculated in two abelian TQFTs related by Galois conjugation are generally different, the entanglement entropy is the same in both theories. Note that the linear behavior of the $S$ matrix under Galois conjugation plays a crucial role in this result.

Before briefly exploring implications of these results for non-Abelian theories, let us note that we may explicitly compute the MEE following from \eqref{qlredden}. For simplicity, focussing on the 2-link case, we obtain (see App. \ref{AppendixA} for details)
\be\label{vNab2}
S_{\rm vN}(\mathcal{L}^2)=\text{ln} \bigg (\frac{\text{det}(K)}{\text{gcd}(\ell_{12},n_1)\text{gcd}(\ell_{12},n_2)\cdots\text{gcd}(\ell_{12},n_N)} \bigg )~,
\ee
where the $n_i$ are the diagonal elements of the Smith normal form, $K_s$, and are therefore the ranks of the individual factors that make up the Abelian fusion group, $\CA$.\foot{As a consistency check, note that for a product TQFT, where $K$ itself is a diagonal matrix, the entanglement entropy becomes the sum of the entanglement entropies of the individual theories.} Note that \eqref{vNab2} shows a manifest symmetry under $\ell_{12}\to\ell_{12}+m\det(K)$ for any integer $m$ since
\begin{equation}
{\rm gcd}(\ell_{12}+m\det(K),n_i)={\rm gcd}(\ell_{12},n_i)~, \ \ \ \forall n_i~.
\end{equation}

Actually, this same periodicity is already visible in \eqref{Lnstate}. Indeed, from \eqref{SandTfromTheta} and \eqref{Bab}, we have (up to an unimportant normalization)
\begin{equation}
S_{\vec j_i,\vec j_2}\sim\exp\left(2\pi i\vec j_1K^{-1}\vec j_2\right)=\left(2\pi i{\vec j_1\tilde K\vec j_2\over\det(K)}\right)~,
\end{equation}
where $\tilde K$ is the integer-valued adjugate matrix. Therefore, taking $\ell_{ik}\to\ell_{ik}+m\det(K)$ in \eqref{Lnstate} leaves $|\CL^n\rangle$ and the MEE invariant. It is also straightforward to use \eqref{SandTfromTheta} and \eqref{Bab} to establish that \eqref{Lnstate} and the MEE are invariant under arbitrary integer shifts of the linking numbers by the MTC conductor, $\ell_{ik}\to\ell_{ik}+mN_0$.

The upshot of the above discussion is that Galois transformations of abelian theories preserve the multiboundary entanglement entropy. However, this result hinges on the fact that, for any link complement, only the $S$ matrix enters the computation. Moreover, the $S$ matrix has linear transformation properties under the Galois action. For more general TQFTs we therefore expect a more subtle situation. For example, we expect the $T$ matrix to play a more prominent role (i.e., that it will not just appear through $S$), and we have seen that both it and the entanglement entropy are sensitive to the conductor.\footnote{In the theories described above, the period of the link state and the MEE can be finer than the conductor (although these quantities are also periodic modulo the conductor). For example, in $\mathds{Z}_2$ TQFT, we have $N_0=4$, but $\det(K)=2$. The reason for this difference is that the link state and MEE depend on $T$ only through the (unnormalized) $S$ matrix.}

\newsec{Non-Abelian TQFTs}
In this section, we generalize the abelian TQFT discussion to non-abelian theories. Before proceeding, it is worth considering what such a generalization should look like. To that end, let us make a few comments:
\begin{itemize}
\item{In the abelian case, the density matrix can be written exclusively in terms of the $S$-matrix. This simplification is due to the fact that abelian theories are only sensitive to linking number. On the other hand, non-abelian theories compute more complicated invariants: the Jones polynomial, the HOMFLY polynomials, and infinitely many generalizations. Therefore, a non-abelian generalization of our discussion should depend on finer details of the topology of $\CM_{\CL}$. In the broadest terms, a result of Thurston \cite{thurston1982three}, guarantees that $\CM_{\CL}$ can be either a torus link complement, a hyperbolic link complement, or a satellite link complement.\footnote{Torus links are links that can be drawn on the surface of a $T^2$ without self-intersection. Hyperbolic links are links whose complements admit complete hyperbolic metrics. By Thurston's results, satellite links are what remain (we will briefly encounter these links in App. \ref{AppendixB}).} Torus links are naturally in correspondence with words that can be built out of $S$ and $T$ (the generators of the mapping class group of $T^2$). Therefore, this reasoning points to studying a generalization of the abelian result to torus links.}
\item{Another reason to study torus links when searching for a non-abelian generalization of Sec. \ref{abeliansec} comes from the results in \cite{mignard2017modular}. There the authors showed that there are Dijkgraaf-Witten theories with gauge group $\mathds{Z}_q\rtimes_n\mathds{Z}_p$ such that the Galois group acts non-trivially on the corresponding MTCs but leaves the $S$ and $T$ matrices invariant. As a result, torus knot complements are natural places to look to find invariance of the entanglement entropy along Galois orbits. Moreover, recent work in \cite{Bonderson_2019,wen2019distinguish} suggests that hyperbolic link invariants can potentially be used to distinguish MTCs in a gauge-invariant manner. Indeed, it is easy to check that the entanglement entropy of one of the simplest non-abelian TQFTs, $su(2)_k$ Chern-Simons theory, is generically non-invariant along the corresponding Galois orbits when tracing out one of the links of the hyperbolic Whitehead link complement (see App. \ref{AppendixB} for details). Since Whitehead is one of the simplest hyperbolic links (it has linking number zero and is built from two unknots), this result suggests that changes in the entanglement entropy along Galois orbits of theories on hyperbolic link complements is more ubiquitous. Similar comments apply to satellite link complements (see App. \ref{AppendixB} for details).\footnote{Although we suspect that there could be interesting generalizations of our work below to some subclasses of these links as well. For example, a natural set of satellite links to examine are connected sums of torus links.}}
\item{On the other hand, we do not expect all torus link complements to give rise to invariant entanglement entropy along Galois orbits. Indeed, we generally expect the non-Abelian density matrix to depend explicitly on $T$ and not just on $S$. As a result, we expect the topology of the torus link complement to be sensitive to the conductor of the MTC.}
\end{itemize}

To better understand how to proceed, we review torus links in the next subsection. We then revisit the linear transformation properties of $S$ that hold in Abelian and non-Abelian TQFTs alike and use them to identify a canonical class of torus links that give rise to invariant entanglement entropy along Galois orbits of general MTCs.

\subsec{Torus links and canonical words}
Let us recall some useful aspects of torus links. As discussed in previous sections, these links can be drawn on the surface of a $T^2$ without self intersection. They are classified by two integers, $(m,n)$, corresponding to the basis of 1-cycles wrapped by the links. In particular, $m$ corresponds to the number of times the link wraps the longitude of the torus and $n$ corresponds to the number of times the link wraps the meridian.\footnote{An invariant definition of these cycles can be given by imagining filling in the torus to obtain a solid torus. In the solid torus, the meridian becomes contractible while the longitude does not.}

The links may be characterized by the components that make it up. In particular, we have
\begin{equation}\label{linkchar}
\nu(m,n)=\text{gcd}(m,n)~, \ \ \ \ell(m,n)={mn\over{\rm gcd}(m,n)^2}~,
\end{equation}
where $\nu(n,m)$ is the number of components that make up the link (note that for ${\rm gcd}(n,m)=1$, the link is a knot), and $\ell(m,n)$ is the linking number between any two component knots in the link (this is an invariant for any pairs of knots in the link). The knots that make up the link are of type $(m/\nu(m,n),n/\nu(m,n))$. For example, $(2,2)$ is the Hopf link, with $n_L=2$ and $\ell=1$. This link is made up of two $(1,1)$ unknots.

One crucial aspect of our discussion below is that the entanglement entropy arising in torus links does not depend on the number of knots we trace out \cite{Balasubramanian2018EntanglementEA}. More precisely, if our link consists of $\nu(m,n)\ge2$ knots, the entanglement entropy is independent of the number of links, $1\le r\le \nu(m,n)-1$, we trace out.

In order to understand which torus link complements give rise to invariant entanglement entropy under Galois conjugation, it is useful to revisit the Galois transformation properties of the $S$ matrix in \eqref{SandTGalois}. For each element, $Q$, of the Galois group of the modular data, these signed permutations can be generated by\footnote{These matrices were constructed for cases with $C=\mathds{1}$ in \cite{Coste:1999yc} and more generally in \cite{bantay2003kernel,dong2015congruence}.}
\begin{equation}\label{Gmatrices}
G_{\sigma_Q}=\varphi^{2Q+P}S^{-1}T^QST^PST^Q~, \ \ \ Q\cdot P=1\ ({\rm mod}\ N)~,
\end{equation}
where $N$ is the (generalization of the) conductor described around \eqref{bantayres}. In other words, the Galois transformation of the $S$-matrix is
\begin{equation}\label{Sgal2}
Q(S)=G_{\sigma_Q}^{-1}S=SG_{\sigma_Q}~.
\end{equation}
From the perspective of the Galois group, the string of $S$ and $T$ matrices in \eqref{Gmatrices} form a set of \lq\lq canonical" words: the $G_{\sigma_Q}$ are invariant under the Galois group, since all matrix elements are in the set $\left\{-1,0,1\right\}\subset\mathds{Q}$.

Given each $G_{\sigma_Q}$, it is natural that there should be an infinite number of associated torus link complements that give rise to Galois invariant entanglement entropy for non-abelian TQFTs defined on these spaces.\footnote{The reason we expect an infinite number of torus link complements for each $G_{\sigma_Q}$ is that, for each torus knot, we can construct links with arbitrary numbers of these knots.} In the next subsection, we will argue that the complements of torus links of type $(M,MQ)$, with $M\in\mathbb{Z}$ and ${\rm gcd}(Q,N_0)=1$, are precisely such a set of 3-manifolds (recall that $N_0$ is the MTC conductor defined in \eqref{MTCcond}). In what follows, we will refer to these spaces as $\CM_{\CL^{(M,MQ)}}$.

\subsection{Galois invariance of the entanglement entropy on $\CM_{\CL^{(M,MQ)}}$}
\label{sec:(m,mn)}
We begin by deriving an explicit expression for the link invariant of an $(M,MQ)$ torus link from the MTC data. From \eqref{linkchar}, we see that this is an $M$ component link in which the number of braidings between any two knots is $2Q$ (which follows from the mutual linking number, $Q$). Let us look at the braids between the knots labelled by $j_i$ and $j_k$. The $2Q$ braids between these two knots are represented by the operators $(R_{j_i,j_k}R_{j_k,j_i})^Q$ (see Fig. \ref{2qbraids}).

\begin{figure}[h!]   
\centering
\begin{tikzpicture}[x=0.75pt,y=0.75pt,yscale=-1,xscale=1]

\draw    (241.87,99.87) .. controls (243.87,118.87) and (240.82,122.87) .. (221.82,123.87) ;

\draw    (213.82,99.07) .. controls (212.82,111.07) and (216.17,117.33) .. (226.17,120.33) ;

\draw    (240.27,167.47) .. controls (241.87,161.07) and (242.67,153.47) .. (233.47,148.07) ;

\draw    (222.27,148.93) .. controls (235.47,149.13) and (251.02,127.6) .. (233.82,124.6) ;

\draw    (227.42,144.67) .. controls (221.02,137.33) and (203.42,127.07) .. (221.82,123.87) ;

\draw    (213.07,167.47) .. controls (212.27,159.07) and (212.27,152.13) .. (222.27,148.93) ;

\draw  [dash pattern={on 0.84pt off 2.51pt}]  (226.8,173.4) -- (226.87,185.07) ;

\draw    (239.87,231.87) .. controls (240.67,225.07) and (243.47,221.67) .. (233.07,211.67) ;

\draw    (221.87,212.53) .. controls (235.07,212.73) and (242.67,193.47) .. (241.07,187.87) ;

\draw    (227.02,208.27) .. controls (221.07,205.87) and (212.27,197.07) .. (213.07,187.07) ;

\draw    (215.47,230.27) .. controls (212.27,222.27) and (211.87,215.73) .. (221.87,212.53) ;

\draw   (203.07,108.67) .. controls (198.4,108.67) and (196.07,111) .. (196.07,115.67) -- (196.07,156.85) .. controls (196.07,163.52) and (193.74,166.85) .. (189.07,166.85) .. controls (193.74,166.85) and (196.07,170.18) .. (196.07,176.85)(196.07,173.85) -- (196.07,216.87) .. controls (196.07,221.54) and (198.4,223.87) .. (203.07,223.87) ;
\draw    (269.56,165.81) -- (279.89,165.81) ;

\draw    (269.67,169.5) -- (279.89,169.48) ;

\draw    (393.58,101.67) -- (393.58,230.17) ;

\draw    (424.08,101.67) -- (424.08,230.17) ;

\draw (135.5,165.5) node  [align=left] {2Q braidings};
\draw (216,86) node   {$j_{i}$};
\draw (394,88) node   {$j_{i}$};
\draw (395,238) node   {$j_{i}$};
\draw (214,239) node   {$j_{i}$};
\draw (241,240) node   {$j_{k}$};
\draw (243,87) node   {$j_{k}$};
\draw (426,88) node   {$j_{k}$};
\draw (425,239) node   {$j_{k}$};
\draw (337,168) node   {$( R_{j_{i} ,j_{k}} R_{j_{k} ,j_{i}})^{Q}$};

\end{tikzpicture}
\caption{The action of $(R_{j_i,j_k}R_{j_k,j_i})^Q$ on strands of the knots labeled by $j_i$ and $j_k$.}
\label{2qbraids}
\end{figure}

The total invariant can then be found by computing the following quantum trace:
\be\label{qtrace}
\widetilde{\text{Tr}}\Big (\prod_{j_i,j_k} (R_{j_i,j_k}R_{j_k,j_i})^Q \Big)~.
\ee
The operator within the trace acts on the fusion space $V_{j_1,\cdots,j_M}^{j_1,\cdots,j_M}$. In order to compute the quantum trace, we need to specify the operator's action on the fusion space, $V_{j_1,\cdots,j_M}^c$. Since we have 
\be
V_{j_1,\cdots,j_M}^c=\sum_{a_1,\cdots,a_{m-2}} V_{j_1,j_2}^{a_1} \otimes V_{a_1, j_3}^{a_2} \otimes \cdots \otimes V_{a_{m-2},j_M}^c~,
\ee
we can write the operators $\prod_{j_i,j_k} (R_{j_i,j_k}R_{j_k,j_i})^Q$ acting on $V_{j_1,...,j_M}^c$ as
\bea
\sum_{a_1,\cdots,a_{M-2}} (R_{j_1,j_2}^{a_1}R_{j_2,j_1}^{a_1})^Q  (R_{a_1,j_3}^{a_2}R_{j_3,a_1}^{a_2})^Q \cdots (R_{a_{M-2},j_m}^{c}R_{j_M,a_{M-2}}^{c})^Q~.
\eea

We can now evaluate \eqref{qtrace} to obtain 
\bea
&& \widetilde{\text{Tr}}\Big (\prod_{j_i,j_k} (R_{j_i,j_k}R_{j_k,j_i})^Q \Big)= \sum_c d_c \sum_{a_1,\cdots,a_{M-2}} (R_{j_1,j_2}^{a_1}R_{j_2,j_1}^{a_1})^Q  (R_{a_1,j_3}^{a_2}R_{j_3,a_1}^{a_2})^Q \dots (R_{a_{M-2},j_M}^{c}R_{j_M,a_{M-2}}^{c})^Q \nonumber \\ 
&=&  \sum_c d_c \sum_{a_1,\cdots,a_{M-2}}  \text{Tr}\bigg (\bigg (\frac{\theta(a_1)}{\theta(j_1)\theta(j_2)} \bigg )^Q \text{id}_{V_{j_1,j_2}^{a_1}} \otimes  \bigg (\frac{\theta(a_2)}{\theta(a_1)\theta(j_3)} \bigg )^Q \text{id}_{V_{a_1,j_3}^{a_2}}
\otimes \cdots \nonumber \\
&\cdots& \otimes  \bigg (\frac{\theta(c)}{\theta(a_{M-2})\theta(j_M)} \bigg )^Q \text{id}_{V_{a_{M-2},j_M}^{c}} \bigg ) \\
&=& \sum_c d_c \sum_{a_1,\cdots,a_{M-2}} \bigg (\frac{\theta(c)}{\theta(j_1)\theta(j_2)\cdots\theta(j_M)} \bigg )^Q  \text{Tr}\Big (\text{id}_{V_{j_1,j_2}^{a_1}}  \otimes \text{id}_{V_{a_1,j_3}^{a_2}} \otimes \cdots \otimes \text{id}_{V_{a_{M-2},j_M}^{c}} \Big) \nonumber\\
&=&  \sum_c d_c \sum_{a_1,\cdots,a_{M-2}} \bigg (\frac{\theta(c)}{\theta(j_1)\theta(j_2)\cdots\theta(j_M)} \bigg )^Q  N_{j_1,j_2}^{a_1} N_{a_1,j_3}^{a_2} \cdots N_{a_{M-2},j_M}^{c}\nonumber~.
\eea
Since the framing factors, $\theta(j_i)$, can be removed using local unitaries acting on the respective Hilbert spaces, we can ignore them. Using the Verlinde formula \eqref{Verlinde}, we can simplify the above expression to get (up to framing factors we drop)
\be
 \widetilde{\text{Tr}}\Big (\prod_{j_i,j_k} (R_{j_i,j_k}R_{j_k,j_i})^Q \Big)=\sum_{b_{M-1}} \frac{(ST^QS)_{0b_{M-1}}}{S_{00}} \frac{S_{b_{M-1} j_1}S_{b_{M-1} j_2}\cdots S_{b_{M-1}j_M}}{S_{0b_{M-1}}^{M-1}}~.
\ee
Hence, the link state for an $(M,MQ)$ link is given by
\bea
\ket{\CL^{(M,MQ)}}&=&\sum_{j_1,\cdots,j_M} \sum_{b_{M-1}} \frac{(ST^QS)_{0b_{M-1}}}{S_{00}} \frac{S_{b_{M-1} j_1}S_{b_{M-1} j_2}\cdots S_{b_{M-1}j_M}}{S_{0b_{M-1}}^{M-1}}  \ket{j_1,\cdots,j_M}\nonumber \\
&=& \sum_{b_{M-1}}  \frac{(ST^QS)_{0b_{M-1}}}{S_{00}S_{0b_{M-1}}^{M-1}}  \ket{b_{M-1},\cdots,b_{M-1}}~.
\eea
From this data we can compute the eigenvalues of the unnormalized reduced density matrix. They are independent of the number of Hilbert spaces we trace over and are given by
\be
\label{(m,mn)eigen}
\Lambda_{\ell} =\left|\frac{(ST^QS)_{0l}}{S_{00}S_{0l}^{M-1}}\right|^2~.
\ee

Let us now suppose that $Q\in\mathds{Z}_{N}^{\times}$ is a Galois group element for the modular data of the MTC (i.e., we have ${\rm gcd}(Q,N)=1$). The resulting entanglement entropy turns out to be constant along Galois orbits due to the Galois invariance of $G_{\sigma_P}$ in \eqref{Gmatrices} with $P\cdot Q=1\ ({\rm mod}\ N)$. To understand this statement, let us consider the action of $G_{\sigma_P}$ on $S$ 
\bea
\label{stbs}
SG_{\sigma_P}&=&\varphi^{2P+Q}T^PST^QST^P~, \ \ \ Q\cdot P=1\ ({\rm mod}\ N)~,
\eea
which implies that
\begin{equation}\label{Srel2}
ST^QS=\varphi^{-(2P+Q)} T^{-P}SG_{\sigma_P}T^{-P}~.
\end{equation}
Taking $\lambda_{\ell}=S_{00}^2\Lambda_{\ell}$, we then have
\bea\label{lambdader}
\lambda_{\ell}=\left|\frac{(ST^QS)_{0l}}{S_{0l}^{M-1}}\right|^2&=&  \frac{(ST^QS)_{0\ell}}{S_{0\ell}^{M-1}}\frac{(S^{*}T^{*Q}S^{*})_{0\ell}}{{S^{*}_{0\ell}}^{M-1}}       \nonumber \\
&=& \frac{(T^{-P}SG_{\sigma_P}T^{-P})_{0\ell}}{S_{0\ell}^{M-1}}\frac{({T^{*}}^{-P}S^{*}G^{*}_{\sigma_P}{T^{*}}^{-P})_{0\ell}}{{{S^{*}_{0\ell}}^{M-1}}} \hspace{0.7cm}\nonumber \\ \nonumber
&=& \sum_i \frac{S_{0i}(G_{\sigma_P})_{i\ell}T^{-P}_{\ell\ell}}{S_{0\ell}^{M-1}}\sum_j \frac{S_{0j}^*(G_{\sigma_P}^*)_{j\ell}T^{*-P}_{\ell\ell}}{S_{0\ell}^{*M-1}}\\&=&\sum_i \frac{S_{0i}(G_{\sigma_P})_{i\ell}}{S_{0\ell}^{M-1}}\sum_j \frac{S_{0j}^*(G_{\sigma_P})_{j\ell}}{S_{0\ell}^{*M-1}}~.
\eea
In going between the first and second lines we use \eqref{Srel2}, and we use $T_{00}=1$ in going between the second and third lines. Recalling that $G_{\sigma_P}$ induces a signed permutation, we have
\begin{equation}
\lambda_{\ell}= \frac{S_{0\sigma_P(\ell)}}{S_{0\ell}^{M-1}}\frac{S_{0\sigma_P(\ell)}^*}{S_{0\ell}^{*M-1}}~.\
\end{equation}
Clearly, this quantity transforms as a permutation under Galois conjugation by the element $r\in\mathds{Z}_{N}^{\times}$
\be
\lambda_{\ell}\xrightarrow{{r\in\mathds{Z}_N^{\times}}}\frac{S_{0\sigma_r(\sigma_P(\ell))}}{S_{0\sigma_r(\ell)}^{M-1}}\frac{S^*_{0\sigma_r(\sigma_P(\ell))}}{S_{0\sigma_r(\ell)}^{*M-1}}=\frac{S_{0\sigma_P(\sigma_r(\ell))}}{S_{0\sigma_r(\ell)}^{M-1}}\frac{S^*_{0\sigma_P(\sigma_r(\ell))}}{S_{0\sigma_r(\ell)}^{*M-1}}=\lambda_{\sigma_r(\ell)}~,
\ee
where, in the first equality, we used the fact that the Galois group is Abelian.

As a result, we see that the eigenvalues of the normalized reduced density matrix
\begin{equation}
\widehat{\lambda_{\ell}}={\lambda_{\ell}\over\sum_i\lambda_i}~,
\end{equation}
are permuted under the Galois action. Therefore, after tracing out any (proper) subset of links on the 3-manifold $\CM_{\CL^{(M,MQ)}}$ with ${\rm gcd}(Q,N)=1$, the von Neumann and R\'enyi entanglement entropies do not change under Galois conjugation of a TQFT defined on this space. 

In fact, we can prove a stronger statement. Indeed, we have proven a result in terms of the conductor, $N=fN_0$ (where $f$ is a positive integer dividing twelve). The natural conductor in TQFT is $N_0$. In particular, let us consider $Q$ such that ${\rm gcd}(Q,N_0)=1$. If ${\rm gcd}(Q,f)=1$, then we have ${\rm gcd}(Q,N)=1$, and we are back to the discussion above. On the other hand, suppose ${\rm gcd}(Q,f)\ne1$. In this case, we can always take positive integers, $f_{1,2}$, such that $f=f_1f_2$, where ${\rm gcd}(Q,f_2)={\rm gcd}(f_1,f_2)=1$, and all prime factors of $f_1$ divide $Q$ (of course, it may be that $f_2=1$). By construction, we must have ${\rm gcd}(N_0,f_1)=1$. Now, consider the integer
\begin{equation}
Q'=Q+N_0\cdot f_2~.
\end{equation}
Clearly, we have that ${\rm gcd}(Q', N_0)={\rm gcd}(Q',f_2)={\rm gcd}(Q',f_1)=1$. As a result, ${\rm gcd}(Q',f)={\rm gcd}(Q',fN_0)={\rm gcd}(Q',N)=1$. Now, consider the signed permutation matrix
\begin{equation}
G_{\sigma_{P'}}=\varphi^{2P'+Q'}S^{-1}T^{P'}ST^{Q'}ST^{P'}~, \ \ \ Q'\cdot P'=1\ ({\rm mod}\ N)~.
\end{equation}
From the definition of the MTC conductor \eqref{MTCcond}, we see that $T^{Q'}=T^Q$ and so
\begin{equation}
G_{\sigma_{P'}}=\varphi^{2P'+Q'}S^{-1}T^{P'}ST^{Q}ST^{P'}~.
\end{equation}
Following the logic beginning in \eqref{lambdader}, we find the following result:

\bigskip\noindent
{\bf Theorem 1:} The TQFT MEE (and also the associated R\'enyi entropies) obtained by tracing out the Hilbert subspaces associated with any (proper) subset of linking boundary tori on the 3-manifold, $\CM_{\CL^{(M,MQ)}}$, with $Q$ co-prime to the MTC conductor, $N_0$, are invariant under the action of the TQFT Galois group. Implicit in this discussion is the assumption that the non-unitary theories that arise lie on the Galois orbit of at least one unitary theory.\footnote{In particular, the Hilbert space in the statement of the theorem refers to the Hilbert space of a unitary member of this orbit.} Note that, by a modular transformation, the same results apply to $\CM_{\CL^{(MQ,M)}}$.
\bigskip

In the next section we will introduce knot operators. As we will see, properties of these operators combined with the results of this section lead to a vast generalization of Theorem~1 in the case of non-Abelian Chern-Simons theories and their Galois partners.

\subsection{Galois transformations, entanglement entropy, and more general Torus links}
\label{sec:ToruslinksS3}
To find a more general class of link complements giving rise to invariant entanglement entropy along TQFT Galois orbits, it is useful to introduce the concept of knot operators. Using these operators, it is a relatively simple matter to find link invariants for general torus links \cite{labastida1991knot,isidro1993polynomials,stevan2010chern}. The basic idea is to decompose a 3-manifold, $M$, containing Wilson lines by gluing two solid tori, $U_1$ and $U_2$, at their $T^2$ boundaries such that no Wilson line is cut (in this sense we consider \lq\lq local" Wilson lines). The set of manifolds which can be obtained from gluing two solid tori with a boundary homeomorphism given by an element of $SL(2,\mathds{Z})$ are called Lens spaces. For $\mathds{1} \in SL(2,\mathds{Z})$, we get the manifold $S^2 \times S^1$ and for the $S$ matrix we get $S^3$.

The expectation value of the Wilson lines in $M$ can be recast as an inner product of states in a Hilbert space, where the states are found by performing a path integral over the two solid tori. In this formalism, the knot invariant of an $(M,N)$ torus knot is given by the expectation value
\be 
\label{expectationvalue_knotoperator}
\langle W^{(M,N)}_j \rangle _{S^3}= \frac{\bra{0}SW^{(M,N)}_j\ket{0}}{\bra{0}S\ket{0}}~.
\ee
The vector $\ket{0}$ represents the empty solid torus, $U_1$. The action of $W^{(M,N)}_j$ on this state creates the $(M,N)$ torus knot in representation $j$ on its $T^2$ boundary. Applying an $S$ transformation at the torus boundary and gluing in the other solid torus, $U_2$, gives the expectation value of the knot in $S^3.$\footnote{The denominator, $\bra{0}S\ket{0}$, is a normalization factor.}

For Chern-Simons theory with an arbitrary simple gauge group $G$ at level $k$, the action of the torus knot operator, $W_j^{(M,N)}$,  on a state is given by \cite{Labastida_1996}
\be
\label{Knotoperator}
W^{(M,N)}_j \ket{p}= \sum_{ \ell \in \Lambda_j} \exp\left({2\pi i \frac{MN}{\psi^2(2yk+g^{\vee})}\ell^2 + 4\pi i \frac{N}{\psi^2(2yk+g^{\vee})} (p \cdot \ell)}\right) \ket{p+M\ell}~.
\ee
Here $\Lambda_j$ is the set of weights of the irreducible representation $V_{j}$, $y$ is the Dynkin index of the fundamental representation, $\psi^2$ is the length squared of the longest simple root, $k$ is the level, $g^{\vee}$ is the dual Coxeter number, and $p\in\Lambda_W$ is a vector in the weight lattice. For example, the $W^{(1,0)}$ torus knot operator acts as
\be
W^{(1,0)}_j \ket{p}= \sum_{ \ell \in \Lambda_j} \ket{p+\ell}~.
\ee
In terms of fusion matrices, we have
\be
\label{W(1,0)andN}
W_j^{(1,0)} \ket{p}= \sum_{\ell} N_{jp}^{\ell} \ket{\ell}~.
\ee

For finite $k$, the set of states, $\ket{p+M\ell}$, that arise in \eqref{Knotoperator} are subject to relations such that they lie within the class of integrable representations at level $k$. For example, in the case of $su(2)_k$ Chern-Simons theory on $T^2\times\mathds{R}$ (where space is a $T^2$), the states of the Hilbert space are given in terms of combinations of theta functions. These states are subject to the identifications $\ket{-\ell}=-\ket{\ell}$ and $\ket{\ell}=\ket{\ell-2(k+2)}$. The first identification follows form a Weyl reflection, and the second identification follows from a periodicity property of the theta functions involving shifts by the simple root. Using these identifications, we can always reduce the sum in \eqref{Knotoperator} to a sum over states corresponding to the integrable representations. Similar comments apply to more general gauge groups (again, only signs appear in the identification of states).\footnote{Here we will get a larger number of Weyl reflections and more complicated periodicity structure for the relevant theta functions (again these shifts are in one-to-one correspondence with the simple roots).}

In what follows, it will be useful for us to understand more carefully how the $T$ matrix can enter general link invariants. The key is to first phrase the Rosso-Jones formula \cite{Rosso:1993vn} in terms of torus knot operators in the large $k$ limit \cite{stevan2010chern}\footnote{We mean that $k$ is large compared to the quantum numbers of the $W_j^{(M,N)}$ knot operator.}
\be 
\label{Knotoperator_fractionaltwist}
\bra{0}W^{(M,N)}_j\ket{0}=\sum_{\ell} C(M)_j^{\ell} T^{\frac{N}{M}}_{\ell,\ell}\langle0|\ell\rangle~,
\ee
where the sum is over some integrable representations, the $C(M)_j^k\in\mathds{Z}$ are independent of $N$, and $T^{\frac{N}{M}}_{\ell,\ell}$ is the fractional twist. For convenience, we have labeled the vacuum as $|0\rangle$.\footnote{However, when substituting \eqref{Knotoperator} into \eqref{Knotoperator_fractionaltwist}, one should take $|0\rangle\to|\rho\rangle$, where $\rho=\sum_i\lambda^{(i)}$ is the sum over the fundamental weights (similar comments apply to all other appearances of $|0\rangle$ below).} Furthermore, for large $k$, the $C(M)_j^{\ell}$ are specified by the so-called Adams operation,\footnote{For example, in thse case of $su(n)$, the Adams operation is defined as follows. Consider the $su(n)$ Schur polynomials, $\chi_j(z_1,\cdots,z_{n-1})$, where $i$ is an irreducible representation of $su(n)$, and raise the $su(n)$ fugacities to the $M^{\rm th}$ power. Writing the result in terms of the Schur polynomials without transforming the fugacities, we obtain the $C(M)_j^{\ell}$
\begin{equation}
\chi_j(z_1^M,\cdots,z_{n-1}^M)=\sum_lC(M)_j^{\ell}\cdot\chi_{\ell}(z_1,\cdots,z_{n-1})~.
\end{equation}
} and $\langle0|\ell\rangle=\delta_{0,\ell}$ \cite{stevan2010chern}. Therefore, at large $k$, $\bra{0}W^{(M,N)}_j\ket{0}=C(M)^0_j$.

For general $k$ (not necessarily large compared to the quantum numbers of the knot operator), the story is more complicated, since some of the $|\ell\rangle$ appearing in \eqref{Knotoperator_fractionaltwist} should be identified with the vacuum representation. For example, as discussed above in $su(2)_k$ CS theory, $-|2k+2\rangle=|0\rangle$. In any case, \eqref{Knotoperator_fractionaltwist} allows us to control the non-linearities arising in the Galois action on $T$.

Another crucial property of the knot operators is that they satisfy fusion rules
\be
W^{(M,N)}_iW^{(M,N)}_j=\sum_{i} N_{ij}^k W^{(M,N)}_k~.
\ee
Hence, they are generalized Verlinde operators. Using this property, we can write the torus link operator for a $Q$-component torus link $(QM,QN)$ in terms of the knot operators as
\be
W^{(QM,QN)}_{j_1,\cdots,j_Q}= N_{j_1,\cdots,j_Q}^{\ell} W^{(M,N)}_{\ell}~.
\ee
Moreover, any torus knot operator can be obtained from the unknot by the action of an $SL(2,\mathds{Z})$ element
\be
\label{Torus_knot_from_unknot}
W^{(M,N)}_j={F^{(M,N)}}^{-1} W^{(1,0)}_j F^{(M,N)}~,
\ee
where $F^{(M,N)}\in SL(2,\mathds{Z})$. This statement is natural since a torus knot can be put on the surface of a torus without self intersections, and we can obtain such a knot from the unknot by a sequence of Dehn twists and $S$ tranformations. 

A straightforward generalization of the argument in \cite{Balasubramanian2018EntanglementEA} shows that the eigenvalues of the reduced density matrix of the $(QM,QN)$ torus link are given by
\be\label{redQMQN}
\Lambda_{\ell} = \frac{1}{S_{0\ell}^{2Q-2}}\left| \sum_i S_{\ell i} \langle W^{(M,N)}_i\rangle _{S^3}\right|^2~.
\ee
We can massage this expression into a more useful form as follows:

\bigskip
\noindent
{\bf Lemma 1:} The eigenvalues of the reduced density matrix of the $(QM,QP)$ torus link are given by
\be
\label{link2}
\Lambda_{\ell} = \frac{1}{(S_{0\ell})^{2Q-2}S_{00}^2} \sum_{i} S_{\ell i} \langle W^{(M,P)}_i \rangle _{S^2 \times S^1} \sum_{j} S_{\ell j} \langle W^{(-P,M)}_j \rangle _{S^2 \times S^1}~.
\ee
{\bf Proof:} See App. \ref{AppendixC}.
\bigskip

\noindent
In this lemma,  $ \langle W^{(M,P)}_i \rangle _{S^2 \times S^1} =\bra{0}W^{(M,P)}_i \ket{0}$ is the knot invariant of the $(M,P)$ knot in $S^2 \times S^1$. As a result, we can write the entanglement entropy of links in $S^3$ as a product of linear combinations of knot invariants in $S^2\times S^1$ with $S$ matrix elements as coefficients. 

We may now make use of the above lemma to gain a better understanding of Galois transformation properties of torus links in Chern-Simons theory. To that end, first consider the special case of eigenvalues of the reduced density matrix for $(MQ,Q)$ torus links described in Sec. \ref{sec:(m,mn)}. Using \eqref{link2}, we have
\be\label{specfinal}
\Lambda_{\ell}=\frac{1}{(S_{0\ell})^{2Q-2}S_{00}^2} \sum_{i} S_{\ell i} \langle W^{(M,1)}_i \rangle _{S^2 \times S^1} \sum_{j} S_{\ell j} \langle W^{(-1,M)}_j \rangle _{S^2 \times S^1}~.
\ee
We may simplify the second summation in \eqref{specfinal}, since
\begin{equation}
\langle W_j^{(-1,M)}\rangle_{S^2\times S^1}=\langle W_j^{(-1,0)}\rangle_{S^2\times S^1}=\delta_{0,j}~.
\end{equation}
The second equality follows from \eqref{W(1,0)andN} and charge conjugation, while the first equality follows from the fact that $T$ acts trivially on the vacuum and so
\be
\label{Lemma1}
 \langle W^{(P,M+AP)}_j \rangle _{S^2 \times S^1}= \langle T^{A}W^{(P,M)}_j T^{-A}\rangle _{S^2\times S^1}~. 
\ee
As a result, the eigenvalues of the reduced density matrix in \eqref{specfinal} simplify to
\be
\label{(qm,q)eigens}
\Lambda_{\ell}=\frac{1}{(S_{0\ell})^{2Q-3}S_{00}^2} \sum_{i} S_{\ell i} \langle W^{(M,1)}_i \rangle _{S^2 \times S^1}~.
\ee

Now, suppose that $M$ is co-prime to the MTC conductor. From Sec. \ref{sec:(m,mn)}, we know that the normalized eigenvalues of the reduced density matrices for such links are permuted under Galois transformations. Hence, Galois conjugating \eqref{(qm,q)eigens} gives
\be   
G_n\left(\Lambda_{\ell} S_{00}^2\right)=\frac{1}{(S_{0\sigma(\ell)})^{2Q-3}} \sum_{i} S_{\sigma(\ell)i} G_n\left(\langle W^{(M,1)}_i \rangle _{S^2 \times S^1}\right)~,
\ee
where $G_n(\cdots)$ denotes the action of the Galois group element corresponding to $P\in \mathds{Z}_{N_0}^{\times}$. From the invertibility of the $S$ matrix, it follows that
\begin{equation}\label{invarG}
G_P\left(\langle W^{(M,1)}_i \rangle _{S^2 \times S^1}\right)=\langle W^{(M,1)}_i \rangle _{S^2 \times S^1}\in\mathds{Q}~,
\end{equation}
and, from \eqref{Knotoperator_fractionaltwist}, we also have 
\begin{equation}
G_P\left(\langle W^{(M,1)}_i \rangle _{S^2 \times S^1}\right)=\langle W^{(M,P)}_i \rangle _{S^2 \times S^1}\in\mathds{Q}~,
\end{equation}
where we have used the fact that the only source of non-rational numbers in \eqref{Knotoperator_fractionaltwist} is from the fractional twists, and we must further assume that ${\rm gcd}(P,M)=1$ in order to have a well-defined Galois action on the fractional twists. Therefore, $\langle W^{(M,P)}_i \rangle _{S^2 \times S^1}$ is invariant under Galois conjugation.

Following the arguments above, we can also show that $\langle W^{(-P,M)}_i \rangle _{S^2 \times S^1}\in\mathds{Q}$ is invariant under Galois conjugation. Hence, it follows from Lemma 1 that the normalized eigenvalues of $(QM,QP)$ link for $M,P$ coprime to the conductor and to themselves are permuted under Galois conjugation. Therefore, we have the following theorem:

\bigskip
\noindent
{\bf Theorem 2:} The Chern-Simons MEE (and associated R\'enyi entropies) obtained by tracing out the Hilbert subspaces associated with any (proper) subset of linking boundary tori on the 3-manifold, $\CM_{\CL^{(QM,QP)}}$, with $M,P$ co-prime to the Chern-Simons conductor, $N_0$, and to each other are invariant under the action of the TQFT Galois group. 
\bigskip
\noindent

For Chern-Simons theories and their Galois conjugates, this result generalizes Theorem 1 in Sec. \ref{sec:(m,mn)}. However, the proof in Sec. \ref{sec:(m,mn)} was obtained directly using the MTC data without referring to a specific realization of the TQFT, while the above proof depends on the realization of the TQFT as a Chern-Simons theory and \eqref{Knotoperator}, which was constructed for simple gauge groups. The authors of \cite{Witten:1988sy,moore1990lectures} conjectured that every 3D TQFT is a Chern-Simons theory with some gauge group. If this conjecture is true, the results in this section might extend to all 3D unitary TQFTs and their Galois conjugates.

\subsection{Example: $su(2)_k$}
\begin{figure}
\centering
\begin{tikzpicture}[x=0.75pt,y=0.75pt,yscale=-1.5,xscale=1.5]

\draw   (140.4,30.73) -- (300.29,30.73) -- (300.29,149.72) -- (140.4,149.72) -- cycle ;
\draw    (160.08,30.33) -- (159.58,149.83) ;

\draw    (140.19,50.62) -- (300.66,50.98) ;

\draw    (140.08,30.5) -- (159.58,50.42) ;

\draw  [fill={rgb, 255:red, 182; green, 51; blue, 51 }  ,fill opacity=1 ] (280.04,109.86) -- (299.79,109.86) -- (299.79,129.72) -- (280.04,129.72) -- cycle ;
\draw  [color={rgb, 255:red, 9; green, 9; blue, 9 }  ,draw opacity=1 ][fill={rgb, 255:red, 24; green, 87; blue, 148 }  ,fill opacity=1 ] (160.07,50.53) -- (279.58,50.53) -- (279.58,149.5) -- (160.07,149.5) -- cycle ;
\draw  [fill={rgb, 255:red, 24; green, 87; blue, 148 }  ,fill opacity=1 ] (279.79,129.57) -- (300.19,129.57) -- (300.19,149.42) -- (279.79,149.42) -- cycle ;
\draw  [fill={rgb, 255:red, 24; green, 87; blue, 148 }  ,fill opacity=1 ] (279.58,50.52) -- (300.23,50.52) -- (300.23,109.64) -- (279.58,109.64) -- cycle ;
\draw    (140.58,70) -- (300.5,69.97) ;

\draw    (140.48,90.24) -- (300.19,90.14) ;

\draw    (140.76,109.95) -- (300.38,109.64) ;

\draw    (140.76,130.24) -- (300.4,129.72) ;

\draw    (179.93,30.85) -- (180.08,150.11) ;

\draw    (200.15,30.85) -- (200.37,149.81) ;

\draw    (220.15,30.85) -- (220.37,149.81) ;

\draw    (240.15,30.85) -- (240.11,149.35) ;

\draw    (260.08,30.83) -- (260.15,150.04) ;

\draw    (279.58,50.73) -- (279.7,30.63) ;

\draw (155.27,37.8) node [scale=0.8]  {$k$};
\draw (145.87,44.17) node [scale=0.8]  {$Q$};
\draw (171.67,39.5) node [scale=1]  {$2$};
\draw (191.67,39.5) node [scale=1]  {$3$};
\draw (211.67,39.5) node [scale=1]  {$4$};
\draw (231.17,39.5) node [scale=1]  {$5$};
\draw (251.17,39.5) node [scale=1]  {$6$};
\draw (271.17,39.5) node [scale=1]  {$7$};
\draw (291.67,39.5) node [scale=1]  {$8$};
\draw (151,59.9) node [scale=1]  {$2$};
\draw (151,79.5) node [scale=1]  {$3$};
\draw (151,99) node [scale=1]  {$4$};
\draw (151,118.5) node [scale=1]  {$5$};
\draw (151,139) node [scale=1]  {$6$};

\end{tikzpicture}
\caption{Sparseness of Galois non-invariance for the MEE of $(2,2Q)$ torus link complements after tracing out one of the boundaries in $su(2)_k$ theories. The $y$-axis corresponds to $Q$ and the $x$-axis corresponds to $k$. Blue squares correspond to theories and topologies with Galois invariant MEE, while the red square does not.}\label{sparse}
\end{figure}
As a concrete example to illustrate the above discussion, consider $su(2)_k$ CS theory. The modular data for this theory is
\be
\label{su(2)_kmodulardata}
S_{ab}=\sqrt{\frac{2}{k+2}} \text{sin}\left( \frac{\pi (a+1)(b+1)}{k+2}\right)~, \ \ \ T_{ab}=\delta_{ab}\exp\left({\frac{2 \pi i a (a+2)}{4(k+2)}}\right)=\delta_{ab}\theta(a)~,
\ee
where $a\in\left\{0,1,\cdots,k\right\}$. In particular, the MTC conductor is generically $N_0=4(k+2)$. In Fig. \ref{sparse}, we present a set of results for the entanglement entropy after tracing one of the knots in $(2,2Q)$ torus links for levels $2\le k\le8$ and $2\le Q\le6$. The results are completely consistent with the above theorems. In fact, we see various \lq\lq accidental" invariances not guaranteed by our theorems.

The simplest Galois non-invariant entanglement entropy occurs in the $(2,10)$ link of the $su(2)_8$ CS theory. Note that the MTC conductor in this case is $N_0=40$, and $M=1$, $P=5$. Clearly, $(P,N_0)=5\ne1$, and so this lack of invariance is consistent with Theorem 2.

\subsection{Torus links in Lens spaces}\label{lens}
Let us briefly comment on the generalization of Theorem 2 to more general Lens spaces. The expectation value of knot operators in a Lens space, $\mathcal{M}_F$, is given by 
\be 
\langle W^{(m,n)}_j \rangle _{F}= \frac{\bra{0}FW^{(m,n)}_j\ket{0}}{\bra{0}F\ket{0}}~,
\ee  
where $F \in SL(2,\mathds{Z})$ is the homeomorphism between the two tori which produces the corresponding Lens space. Following the procedure in Sec. \ref{sec:ToruslinksS3}, the eigenvalues for torus links in a Lens space specified by $F\in SL(2,\mathds{Z})$ is given by 
\be
\Lambda_l = \frac{1}{(S_{0l})^{2Q-2}F_{00}^2} \sum_{i} S_{li} \langle W^{(m,n)}_i \rangle _{S^2 \times S^1} \sum_{j} S_{lj} \langle F W^{(m,n)}_j F^{-1} \rangle _{S^2 \times S^1}~.
\ee
Therefore, the eigenvalues of torus links in a general Lens space can be written as a linear combination of knot invariants in $S^2 \times S^1$ with $S$ matrix elements as coefficients. A sufficient condition for the Galois invariance of entanglement entropy of a torus link $(QM,QN)$ in $\mathcal{M}_F$ is the Galois invariance of knot invariants $\langle W^{(M,N)}_i \rangle _{S^2 \times S^1}$ and $\langle F W^{(M,N)}_j F^{-1} \rangle _{S^2 \times S^1}$.  

\newsec{Conclusions}
We have argued that, in addition to preserving fusion rules (and 1-form symmetries) of TQFTs, Galois conjugation also preserves MEE in broad classes of theories. In particular, we showed that putting any Abelian TQFT on any link complement in $S^3$ and tracing out Hilbert spaces on any subset of the links leads to an invariant MEE along Galois orbits. We then argued that this theorem generalizes to non-Abelian TQFTs living on infinite classes of torus link complements.

The fact that the invariants of the Galois action include both fusion and, on certain torus link complements, MEE is suggestive of a deeper relation between entanglement, fusion, and modular data. In fact, recent work \cite{Shi:2019mlt,Shi:2019ngw} suggests that the entanglement entropy of \cite{kitaev2006topological,Levin:2006zz} can be used to reconstruct the fusion rules and modular data of a TQFT. It would clearly be interesting to understand how MEE fits more precisely into this story.

We conclude with an application and some comments:
\begin{itemize}
\item Our non-Abelian results involve number theory, and it would be interesting to find applications to this field. Here we begin by recalling that, in 300 BC, Euclid found an algorithm for computing the greatest common divisor of two natural numbers (see \cite{knuth2011art} for a modern discussion). In a similar spirit, we can potentially use our Theorem 1 to give a TQFT-based algorithm to check co-primeness of two natural numbers. Indeed, we have seen that, for any TQFT, invariance of the MEE on the $(M,MQ)$ link complement is guaranteed if $Q$ is co-prime to the MTC conductor, i.e. ${\rm gcd}(Q, N_0)=1$. On the other hand, when $Q$ is not co-prime to the MTC conductor, this invariance is not guaranteed.\footnote{Note that in many theories, such as $su(2)_k$ CS theory, there is still \lq\lq accidental" invariance---see Fig. \ref{sparse}.} It would be interesting to try to find an infinite family of TQFTs with infinitely many different conductors that have invariant MEE if and only if $Q$ is co-prime to the conductor. In this case, if we wish to check $(a,b)=1$, we set $Q=a$, $N_0=b$, and check the Galois invariance of the MEE on the $(M, MQ)$ link complement.
\item It would be interesting to understand how more general symmetry structures transform under Galois conjugation. We already know that 1-form symmetry is invariant, but there are larger symmetry structures in TQFT. For example, we have preliminary results on how these transformations affect 2-group symmetry \cite{galsymm}.
\item Anytime one considers symmetry structures, there are corresponding anomalies to contemplate. In fact, the MEE studied in this paper has been related to anomalies in \cite{hung2018linking,zhou20193}. This begs the question of how the Galois action acts on anomalies more generally.
\item Some of the TQFTs discussed here also appear in the context of topological string theory. It would therefore be interesting to understand if there is a notion of a Galois action on topological strings.
\item Our results show that, in some sense, torus link complements are not particularly good at distinguishing Galois conjugates. This result echoes recent work in \cite{mignard2017modular} disproving a conjecture that the modular data (combined with the topological central charge) provide a gauge-invariant means to classify MTCs. It would be interesting to see more generally what MEE has to say about such classifications for non-torus knots (perhaps generalizing or proving conjectures in \cite{Bonderson_2019,wen2019distinguish}).
\item We have only briefly mentioned the entanglement entropy of \cite{kitaev2006topological,Levin:2006zz}. As discussed in Sec. \ref{abeliansec}, it is trivial to show this entropy is also invariant under Galois conjugation in the Abelian case (since it is the square root of the number of anyons).\footnote{Note that, as in our Definition 1, we imagine computing the entanglement entropy in theory $\CT$. We then Galois conjugate $\CT$ to produce $\CT'$. Finally, we compute the entanglement entropy in $\CT'$ and compare the result to that in $\CT$.} For non-Abelian theories it is generally non-invariant, but it also contains less data than the MEE considered here.
\item We have studied the von Neumann entropy for a particular natural state defined by the partition function of the TQFT on the link complement. Are there other interesting states to study?
\item Since many RG flows end in 3D TQFTs, it would be interesting to understand if the Galois action has something to say about 3D QFT away from the topological limit.
\end{itemize}
We hope to return to some of these questions soon.

\ack{We thank A.~Gromov, Z.~Komargodski, A.~Konechny, L.~Li, M.~Mezei, N.~Nekrasov, K.~Ohmori, S.~Ramgoolam, and L.~Tizzano for enlightening discussions. MB thanks the hospitality of the Simons Center for Geometry and Physics where part of this work was completed. MB is partially supported by the Royal Society under the grant, \lq\lq New Constraints and Phenomena in Quantum Field Theory." MB and RR are partially supported by the Royal Society under the grant, \lq\lq New Aspects of Conformal and Topological Field Theories Across Dimensions."}

\newpage
\begin{appendices}
\section{Entanglement entropy of 2-links in abelian TQFTs}
\label{AppendixA}
In the main text, we derived the link state for a 2-link in a general abelian CS theory using the $K$-matrix formalism. Here we will obtain an explicit expression for the entanglement entropy of this state as in \eqref{vNab2}. For the purposes of this computation, it will be useful to choose a particular basis for the lattice, $\mathds{Z}^N/K\mathds{Z}^N$.  

\bigskip
\noindent \textbf{Claim: } The set of vectors  $(a_1,\cdots,a_N)$ where $a_i \in \mathds{Z}_{n_i}, 1\leq i\leq N$,  is a basis set for the lattice $\mathds{Z}^N/UKU^T\mathds{Z}^N$. As a result, these vectors label the anyons (we will call this basis the \lq\lq Smith basis"). Here $U$ and $W$ are matrices which satisfy $K_S=UKW$, where $K_S$ is the Smith normal form of $K$.  

\bigskip
\noindent \textbf{Proof:} Except for the zero vector, every vector of the type  $\vec{a}=(a_1,\cdots,a_N)$ where $a_i \in \mathds{Z}_{n_i}, 1\leq i \leq N,$ satisfies
\be
\vec{a}\neq K_{S}\vec{n}~,
\ee
for any $\vec{n}\in \mathds{Z}^N/K\mathds{Z}^N$. Let U and W be invertible matrices over the integers  such that
\be
K_S=UKW~.
\ee
Then,
\bea
\vec{a}\neq UKW \vec{n}~,\nonumber\\
\vec{a}\neq UKU^T (U^T)^{-1} W \vec{n}~,\\
\vec{a}\neq UKU^T \vec{n}^{'}~,\nonumber
\eea
where $\vec{n}^{'}=(U^T)^{-1} W \vec{n}$. Given that $U$ and $W$ are invertible over the integers, for any $\vec{n}\in \mathds{Z}^N$ we have a unique $\vec{n}^{'}\in \mathds{Z}^N$. Thus,
\be
\vec{a} \neq UKU^T \vec{n}^{'}~,
\ee
for any $\vec{n}^{'}$.

This result means that the above choice of vectors are not linear combinations of columns of $UKU^T$. Since this statement is also true for differences of vectors of the above type, they are all independent and form a basis for the anyons as long as we take the level matrix to be $UKU^T$. Note that the TQFT corresponding to $UKU^{T}$ is the same as that corresponding to $K$, because it corresponds to a change of gauge fields $\vec{A} \rightarrow U^{T} \vec{A}$ where $\vec{A}$ is the vector of gauge fields, $A^{i}$, contained in the action. 

The upshot of the above argument is that the Smith basis can be used to label the anyons as long as we take $UKU^T$ as the level matrix of the theory. Next we will see the implication for the entanglement entropy of the theory.

To that end, the reduced density matrix of a 2-link is given by 
\be
\rho_{\rm red}=\frac{1}{|\CA|^2} \sum_{\vec{j_1},\vec{h_1}} \sum_{\vec{m}} \Big (B(\vec{j}_1, \vec{m}) \Big )^{l_{12}} \Big (B(\vec{h}_1, \vec{m}) \Big )^{-l_{12}} \ket{\vec{j_1}} \bra{\vec{h_1}}~.
\ee
Using \eqref{Bab}, we can write the components of the reduced density matrix ${\rho_{\rm red}}_{\vec{j_1},\vec{h_1}}$ as 
\bea
{\rho_{\rm red}}_{\vec{j_1},\vec{h_1}}&=& \frac{1}{|\CA |^2} \sum_{\vec{m}}  e^{2 \pi i l_{12} (\vec{j_1}-\vec{h_1}) K^{-1} \vec{m}}  \nonumber\\
&=& \frac{1}{|\CA |^2} \sum_{\vec{m}}  e^{2 \pi i l_{12} (\vec{j_1}-\vec{h_1}) K^{-1} \vec{m}} \cdot 1  \nonumber\\
&=& \frac{1}{|\CA |^2} \sum_{\vec{m}}  e^{2 \pi i l_{12} (\vec{j_1}-\vec{h_1}) K^{-1} \vec{m}} \cdot e^{-2 \pi i l_{12} \vec{m} K^{-1} K \vec{\beta}} \\ 
&=& \frac{1}{|\CA |} \delta_{l_{12}(\vec{j_1}-\vec{h_1}),K\vec{\beta}}~,\nonumber
\eea 
for some vector $\vec{\beta} \in \mathds{Z} ^{n}$.

Let us now calculate the $m^{\rm th}$ R\'enyi entropy
\bea
S_m(\mathcal{L}^2)&=&\frac{1}{1-m} \text{ln } \mathrm{tr}(\rho^m) \nonumber\\
&=& \frac{1}{1-m} \mathrm{ln} \bigg (\sum_{\vec{a_1},\vec{a_2},\cdots\vec{a_m}} \rho_{\vec{a_1},\vec{a_2}}\rho_{\vec{a_2},\vec{a_3}}\cdots\rho_{\vec{a_m},\vec{a_1}} \bigg ) 
\eea
\be
= \frac{1}{1-m} \mathrm{ln} \bigg (\frac{1}{|\CA |^m} \sum_{\vec{a_1},\vec{a_2},\cdots\vec{a_m}}  \delta_{l_{12}(\vec{a_1}-\vec{a_2}),K\vec{\beta}}  \delta_{l_{12}(\vec{a_2}-\vec{a_3}),K\vec{\beta}} \cdots  \delta_{l_{12}(\vec{a_m}-\vec{a_1}),K\vec{\beta}} \bigg )~. \nonumber \\
\ee
In order to simplify the above expression, we have to calculate the number of vectors in the basis which satisfy $l_{12} (\vec{a_1}-\vec{a_2})=K \vec{\beta}$ for some $\vec{\beta}$. For simplicity, let us choose the basis to be the Smith basis for which we have to take the level matrix to be $UKU^{T}$. Let us find the number of solutions of $l_{12} \vec{a} = UKU^{T} \vec{\beta}$, where $\vec{a}$ belongs to the Smith basis and $\vec{\beta}$ is an arbitrary vector. This equation is the same as $l_{12} \vec{a} = K_s \vec{\beta}^{'}$, where $\vec{\beta}^{'}= W^{-1} U^{T} \vec{\beta}$. The matrices $U$ and $W$ are uni-modular and satisfy $K_S=UKW$, $K_s$ being the Smith normal form of $K$. This reasoning gives us a set of equations
\be
l_{12} a_1 = n_1 \beta^{'}_1 ; l_{12} a_2=n_2 \beta^{'}_2 ;\cdots; l_{12} a_N=n_N \beta^{'}_N~,
\ee
where $a_i$ and $\beta_i$ are components of $\vec{a}$ and $\vec{\beta}$, respectively, and $n_i$ are the diagonal elements of $K_s$. These quantities can also be written as
\be
a_1 = 0 \text{ mod }\frac{n_1}{\mathrm{gcd}(l_{12},n_1)}~, \ a_2 = 0 \text{ mod }\frac{n_2}{\mathrm{gcd}(l_{12},n_2)}~,\ \cdots~, a_N = 0 \text{ mod }\frac{n_N}{\mathrm{gcd}(l_{12},n_N)}~.
\ee
Since $a_i\in \{0,1,\cdots,n_i-1\}$, the solutions of the above equations can be parametrized as
\be
a_1 =\frac{r_1n_1}{\mathrm{gcd}(l_{12},n_1)}~,\ a_2 = \frac{r_2n_2}{\mathrm{gcd}(l_{12},n_2)}~,\ \cdots~, a_N =\frac{r_Nn_N}{\mathrm{gcd}(l_{12},n_N)}~,
\ee
where $r_i\in \{0,1,\cdots,\mathrm{gcd}(l_{12},n_i)\}$. Hence, the number of $\vec{a}$ which satisfy $l_{12} \vec{a} = K_s \vec{\beta}^{'}$ is given by $\prod_{i=1}^{N}\mathrm{gcd}(l_{12},n_i)$. Similarly, for a given vector, $\vec{a_2}$, in the Smith basis, the number of $\vec{a_1}$ which satisfy $l_{12} (\vec{a_1}-\vec{a_2})=K \vec{\beta}$ for some $\vec{\beta}$ is $\prod_{i=1}^N\mathrm{gcd}(l_{12},n_i)$. Using this result, the $n^{\rm th}$ R\'enyi entropy can be written as
\bea
S_m(\mathcal{L}^2)&=& \frac{1}{1-m} \mathrm{ln} \bigg (\frac{1}{|\CA |^m} \sum_{\vec{a}_1} \delta_{\vec{a_1},\vec{a_1}} (\mathrm{gcd}(l_{12},n_1)\mathrm{gcd}(l_{12},n_2)\cdots\mathrm{gcd}(l_{12},n_N))^{m-1}\bigg ) \nonumber \\
&=& \frac{1}{1-m} \mathrm{ln} \bigg (\frac{1}{|\CA |^{m-1}} ({\rm gcd}(l_{12},n_1){\rm gcd}(l_{12},n_2)\cdots \mathrm{gcd}(l_{12},n_N))^{m-1} \bigg ) \\
&=& \mathrm{ln} \bigg (\frac{\mathrm{det}(K)}{\mathrm{gcd}(l_{12},n_1)\mathrm{gcd}(l_{12},n_2)\cdots\mathrm{gcd}(l_{12},n_N)} \bigg )~.\nonumber
\eea
As a result, the entanglement entropy of a 2-link in a general abelian theory with level matrix $K$ is given by 
\be
S_{\rm vN}(\mathcal{L}^2)=\mathrm{ln} \left(\frac{\mathrm{det}(K)}{\mathrm{gcd}(l_{12},n_1)\mathrm{gcd}(l_{12},n_2)\cdots\mathrm{gcd}(l_{12},n_N)} \right)~.
\ee

\section{Results for hyperbolic and satellite link complements}
\label{AppendixB}
Knots and links are classified into three types: torus, hyperbolic, and satellite. In abelian theories, all three kinds of links have invariant entanglement entropy under the action of the Galois group. Motivated by the special role that the modular generator $S$ plays in this result, we looked at torus links in non-abelian TQFTs and analyzed the behavior of their entanglement entropy under Galois cojugation. Given that an infinite subset of these links have Galois invariant entanglement entropy, it is natural to ask whether similar results hold in the case of hyperbolic and satellite links.    

It turns out that, in general,  the entanglement entropy of hyperbolic links are different in two TQFTs related by Galois conjugation. For example, the Whitehead link is one of the simplest hyperbolic links in the sense of having just two components and minimal hyperbolic volume (for a two cusped hyperbolic manifold). Even for this link the entanglement entropy changes under Galois conjugation. We verify this statement in $su(2)_k$ CS theory for small $k$.

\tikzset{every picture/.style={line width=0.75pt}} 

\begin{figure}
\centering
\begin{tikzpicture}[x=0.75pt,y=0.75pt,yscale=-1,xscale=1]

\draw    (277.3,99.2) .. controls (275.3,9.2) and (341.3,-4.8) .. (353.3,70.2) ;

\draw    (279.3,113.2) .. controls (292.3,184.2) and (353.3,159.2) .. (356.3,87.2) ;

\draw    (285.3,73.2) .. controls (308.6,86.4) and (317.3,121.2) .. (345.3,117.2) ;

\draw    (300.3,100.2) .. controls (195.3,152.2) and (211.6,47.4) .. (272.3,67.2) ;

\draw    (317.3,89.2) .. controls (427.6,43.4) and (437.3,117.2) .. (358.3,116.2) ;

\draw (290,20) node   {$j_{1}$};
\draw (213,88) node   {$j_{2}$};
\end{tikzpicture}
\caption{Whitehead Link}
\end{figure}
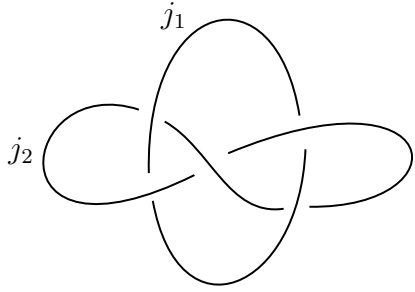

The link state for the Whitehead link in $su(2)_k$ Chern-Simons theory can be found using its link invariant \cite{habiro2000colored,habiro2008unified}
\be
C(j_1,j_2)_{5_1^2}=\sum_{i=0}^{\text{min}(2j_1,2j_2)}q^{-\frac{i(i+3)}{4}}(q^{\frac{1}{2}}-q^{-\frac{1}{2}})^{3i}\frac{[2j_1+i+1]![2j_2+i+1]![i]!}{[2j_1-i]![2j_2-i]![2i+1]!}~,
\ee
where $[x]=\frac{q^{\frac{x}{2}}-q^{\frac{-x}{2}}}{q^{\frac{1}{2}}-q^{-\frac{1}{2}}}$, $[x]!=[x][x-1]\cdots[1]$ and $q=e^{\frac{2\pi i}{k+2}}$.  
In $su(2)_3$ Chern-Simons theory the entanglement entropy and its Galois conjugations are given by
\bea
{5_1^2}_{su(2)_3}&=&\begin{pmatrix}
0.762866  & 0.237134 & 0 &0 \\
0.925325  & 0.0746746 & 0 &0 \\
0.925325  & 0.0746746 & 0 &0 \\
0.762866  & 0.237134 & 0 &0 \\
\end{pmatrix}~.
\eea
The columns are labelled by the integrable representations, $0,1,2,3$,\footnote{We label representations by the Dynkin label (i.e., twice the spin).} and the rows are labelled by the Galois conjugations corresponding to $1,2,3,4\in \mathds{Z}_5^{\times}$. 

Let us now consider satellite links. Examples of such links include connected sums of links. If a link $\mathcal{L}$ is  a connected sum of $\CL _1$ and $\CL _2$, then their invariants satisfy\cite{witten1989quantum}
\be
\label{connectedsuminv}
C_{\CL} \cdot C_{O}(i)= C_{\CL _1} \cdot C_{\CL _2}~,
\ee
where the links $\CL$, $\CL _1$, and $\CL _2$ are to be labelled in a consistent manner. $C_{O}(i)$ is the knot invariant of the unknot labelled by $i$ in $S^3$ and $i$ is the label of the knot which is cut to obtain $\CL _1$ and $\CL _2$ from $\CL$. This implies that the entanglement entropy of most satellite links will change non-trivially under Galois conjugation. For example, let us consider the link which is obtained from a connected sum of the Trefoil knot and the Whitehead link.       
       
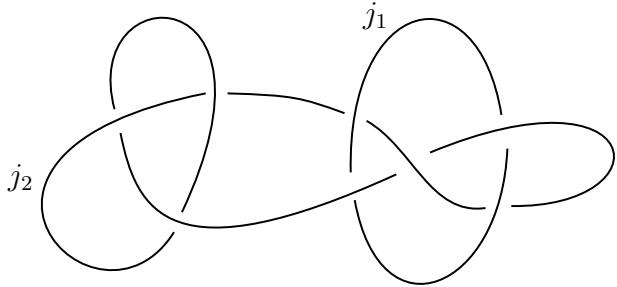
\begin{figure}[h!]
\centering
\begin{tikzpicture}[x=0.75pt,y=0.75pt,yscale=-1,xscale=1]

\draw    (351.3,108.2) .. controls (349.3,18.2) and (415.3,4.2) .. (427.3,79.2) ;

\draw    (353.3,122.2) .. controls (366.3,193.2) and (427.3,168.2) .. (430.3,96.2) ;

\draw    (359.3,82.2) .. controls (382.6,95.4) and (391.3,130.2) .. (419.3,126.2) ;

\draw    (374.3,109.2) .. controls (260.17,161.33) and (243.17,128.33) .. (235.17,88.33) ;

\draw    (391.3,98.2) .. controls (501.6,52.4) and (511.3,126.2) .. (432.3,125.2) ;

\draw    (289.17,68.33) .. controls (322.17,69.33) and (326.17,70.33) .. (348.17,78.33) ;

\draw    (262.17,138.33) .. controls (224.17,199.33) and (125.17,98.33) .. (278.17,68.33) ;

\draw    (232.17,76.33) .. controls (214.17,8.33) and (323.17,7.33) .. (266.17,128.33) ;

\draw (364,29) node   {$j_{1}$};
\draw (185,111) node   {$j_{2}$};
\end{tikzpicture}
\caption{Connected sum of Trefoil and Whitehead Link}
\end{figure}

The knot invariant for the Trefoil knot in $su(2)_k$ Chern-Simons theory is given by \cite{habiro2000colored} \cite{habiro2008unified}
\be
C_{3_1}(j_1)= \sum_{i}^{2j_1} (-1)^i q^{-i(i+3)} (q-q^{-1})^{2i} \frac{[2j_1+i+1]!}{[2j_1-i]!}~,
\ee
where the definitions of $q$ and $[x]$ are the same as above. Using \eqref{connectedsuminv}, the link state and its entanglement entropy can be calculated for the connected sum of Trefoil and Whitehead link. In $su(2)_3$, the eigenvalues of the reduced density matrix for this link and its behaviour under Galois conjugation is given by 
\bea
{8_1^2}_{su(2)_3}&=&\begin{pmatrix}
0.988779  & 0.0112213  & 0 &0 \\
0.972184  &  0.0278156 & 0 &0 \\
0.972184  &  0.0278156 & 0 &0 \\
0.988779  & 0.0112213  & 0 &0 \\
\end{pmatrix}~,
\eea
where the columns are labelled by the integrable representations $0,1,2,3$, and the rows are labelled by the Galois conjugations corresponding to $1,2,3,4\in \mathds{Z}_5^{\times}$.

\tikzset{every picture/.style={line width=0.75pt}} 

\begin{figure}[h!]
\centering
\begin{tikzpicture}[x=0.75pt,y=0.75pt,yscale=-1,xscale=1]

\draw   (140.4,30.73) -- (300.29,30.73) -- (300.29,90.46) -- (140.4,90.46) -- cycle ;
\draw    (160.08,30.33) -- (160.07,90.46) ;

\draw    (140.19,50.62) -- (300.66,50.98) ;


\draw  [color={rgb, 255:red, 9; green, 9; blue, 9 }  ,draw opacity=1 ][fill={rgb, 255:red, 182; green, 51; blue, 51 }  ,fill opacity=1 ] (160.07,50.53) -- (279.58,50.53) -- (279.58,90.46) -- (160.07,90.46) -- cycle ;
\draw  [fill={rgb, 255:red, 182; green, 51; blue, 51 }  ,fill opacity=1 ] (279.58,50.52) -- (300.23,50.52) -- (300.23,90.46) -- (279.58,90.46) -- cycle ;
\draw    (140.58,70) -- (300.5,69.97) ;

\draw    (179.93,30.85) -- (180.1,90.3) ;

\draw    (200.14,30.98) -- (200.25,90.73) ;

\draw    (220.15,30.85) -- (220.33,90.19) ;

\draw    (240.15,30.85) -- (240.25,90.73) ;

\draw    (260.08,30.83) -- (260,90.48) ;

\draw    (279.58,50.73) -- (279.7,30.63) ;

\draw  [fill={rgb, 255:red, 24; green, 87; blue, 148 }  ,fill opacity=1 ] (160.07,50.53) -- (179.82,50.53) -- (179.82,69.95) -- (160.07,69.95) -- cycle ;
\draw  [fill={rgb, 255:red, 24; green, 87; blue, 148 }  ,fill opacity=1 ] (160.07,69.95) -- (179.82,69.95) -- (179.82,90.26) -- (160.07,90.26) -- cycle ;
\draw  [fill={rgb, 255:red, 24; green, 87; blue, 148 }  ,fill opacity=1 ] (200.07,50.63) -- (219.82,50.63) -- (219.82,69.87) -- (200.07,69.87) -- cycle ;
\draw  [fill={rgb, 255:red, 24; green, 87; blue, 148 }  ,fill opacity=1 ] (200.25,70.07) -- (220,70.07) -- (220,90.24) -- (200.25,90.24) -- cycle ;

\draw (150.27,39.8) node [scale=0.8]  {$k$};
\draw (171.67,39.5) node [scale=1]  {$2$};
\draw (191.67,39) node [scale=1]  {$3$};
\draw (211.67,39.5) node [scale=1]  {$4$};
\draw (231.17,39.5) node [scale=1]  {$5$};
\draw (251.17,39.5) node [scale=1]  {$6$};
\draw (271.17,39) node [scale=1]  {$7$};
\draw (291.67,39.5) node [scale=1]  {$8$};
\draw (150.4,60.9) node [scale=1]  {$5^{2}_{1}$};
\draw (151.33,80.83) node [scale=1]  {$8^{2}_{1}$};

\end{tikzpicture}
\caption{The MEEs for the $5_1^2$ and $8_1^2$ links are Galois non-invariant in most of the theories we checked. As in Fig. \ref{sparse}, blue squares correspond to Galois invariant MEE while red squares correspond to Galois non-invariant MEE. In contrast to Fig. \ref{sparse}, there are a lot more red squares.}\label{redfig}
\end{figure}

For a few additional levels, we have checked the Galois conjugation properties of the MEE in Fig. \ref{redfig}. Note that there are many more non-invariant theories in this case than in the torus link case checked in Fig. \ref{sparse}.

\section{Proof of Lemma 1}
\label{AppendixC}
In this appendix, we prove the following Lemma:

\bigskip
\noindent
{\bf Lemma 1:} The eigenvalues of the reduced density matrix of the $(QM,QP)$ torus link are given by
\be
\label{(qm,qn)linkeigenfinal}
\Lambda_{\ell} = \frac{1}{(S_{0\ell})^{2Q-2}S_{00}^2} \sum_{i} S_{\ell i} \langle W^{(M,P)}_i \rangle _{S^2 \times S^1} \sum_{j} S_{\ell j} \langle W^{(-P,M)}_j \rangle _{S^2 \times S^1}
\ee
{\bf Proof:} Using \eqref{redQMQN} and \eqref{expectationvalue_knotoperator}, we have
\bea
\Lambda_{\ell} &=& \frac{1}{S_{0\ell}^{2Q-2}}\left| \sum_i S_{\ell i} \frac{\bra{0} S W^{(M,P)}_i\ket{0}}{\bra{0}S\ket{0}}\right|^2 \nonumber\\
&=& \frac{1}{S_{0\ell}^{2Q-2}}\left| \sum_i S_{\ell i} \frac{\bra{0} S {F^{(M,P)}}^{-1} W^{(1,0)}_i F^{(M,P)} \ket{0}}{\bra{0}S\ket{0}}\right|^2 \ \ \ \text{ (from \eqref{Torus_knot_from_unknot})}\\
&=& \frac{1}{S_{0\ell}^{2Q-2}}\left| \frac{\bra{0} S {F^{(M,P)}}^{-1} \sum_i S_{\ell i} N_i F^{(M,P)} \ket{0}}{\bra{0}S\ket{0}}\right|^2 \ \ \ \ \ \ \  \text{ (from \eqref{W(1,0)andN})}~.\nonumber
\eea
Using \eqref{Verlinde}, we can simplify this expression to obtain
\be
\Lambda_{\ell} =  \frac{1}{(S_{0\ell})^{2Q}S_{00}^2}  \left|(S(F^{(M,P)})^{-1}S^{-1})_{0 \ell^{*}}  (SF^{(M,P)})_{\ell^{*} 0}\right|^2
\ee
Since $S$ and $F^{(M,P)}$ are unitary, we have
\be
\label{simplifiedeigen}
\Lambda_{\ell} = \frac{1}{(S_{0\ell})^{2Q}S_{00}^2} (S(F^{(M,P)})^{-1}S^{-1})_{0 \ell^{*}} (SF^{(M,P)}S^{-1})_{\ell^{*} 0} (SF^{(M,P)})_{\ell^{*} 0} ((F^{(M,P)})^{-1} S^{-1})_{0 \ell^{*}}~.
\ee
To further simplify the above expression, note that, from \eqref{Verlinde}
\bea
\sum_{i} S_{\ell i} \bra{0}W^{(M,P)}_i \ket{0}&=& \sum_i S_{\ell i} ((F^{(M,P)})^{-1}W_i^{(1,0)}F^{(M,P)})_{0 0} \nonumber\\
\label{key1}
&=& \frac{((F^{(M,P)})^{-1}S^{-1})_{0\ell^{*}}  (SF^{(M,P)})_{\ell^{*} 0}}{S_{0\ell}}~,
\eea
and
\bea
\sum_{j} S_{\ell j} \bra{0}W^{(-P,M)}_j \ket{0}&=& \sum_{j} S_{\ell j} \bra{0} S W^{(M,P)}_j S^{-1} \ket{0} \nonumber\\
&=&\sum_j S_{\ell j} (S(F^{(M,P)})^{-1}W_j^{(1,0)}F^{(M,P)}S^{-1})_{0 0} \\
\label{key2}
&=&\frac{(S(F^{(M,P)})^{-1}S^{-1})_{0\ell^{*}}  (SF^{(M,P)}S^{-1})_{\ell^{*} 0}}{S_{0\ell}}\nonumber~.
\eea
Using these equations, we can write the expression for the eigenvalues in \eqref{simplifiedeigen} as
\be
\label{(qm,qn)linkeigenfinal}
\Lambda_{\ell} = \frac{1}{(S_{0\ell})^{2Q-2}S_{00}^2} \sum_{i} S_{\ell i} \langle W^{(M,P)}_i \rangle _{S^2 \times S^1} \sum_{j} S_{\ell j} \langle W^{(-P,M)}_j \rangle _{S^2 \times S^1}~.
\ee
\end{appendices}
{\bf q.e.d.}

\newpage
\bibliography{chetdocbib}

\end{document}